\documentclass[review, preprint]{elsarticle}

\usepackage{framed,multirow}

\usepackage{amssymb}
\usepackage{booktabs}
\usepackage{amstext}
\usepackage{mathptmx}
\usepackage{amsmath}
\usepackage[percent]{overpic}
\usepackage{color}

\usepackage{lineno}

\def\revHC#1{\textcolor{blue}{#1}}

\journal{Journal of Computational Physics}

\begin{document}


\begin{frontmatter}



\title{Turbulence Model Development based on a Novel Method Combining Gene Expression Programming with an Artificial Neural Network}


\author[1,3]{Haochen Li}
\author[4]{Fabian Waschkowski}
\author[1,2,3]{Yaomin Zhao\corref{cor1}}
\ead{yaomin.zhao@pku.edu.cn}
\author[4]{Richard D. Sandberg}
\cortext[cor1]{Corresponding author}
\address[1]{HEDPS, Center for Applied Physics and Technology, and College of Engineering, Peking University, Beijing 100871, China}
\address[2]{Joint Laboratory of Marine Hydrodynamics and Ocean Engineering, Pilot National Laboratory for Marine Science and Technology, Qingdao, Shandong 266237, China}
\address[3]{State Key Laboratory for Turbulence and Complex Systems, College of Engineering, Peking University, Beijing 100871, China}
\address[4]{Department of Mechanical Engineering, University of Melbourne, VIC 3010, Australia}

\begin{abstract}
Data-driven methods are widely used to develop physical models, but there still exist limitations that affect their performance, generalizability and robustness.
By combining gene expression programming (GEP) with artificial neural network (ANN), we propose a novel method for symbolic regression called the gene expression programming neural network (GEPNN).
In this method, candidate expressions generated by evolutionary algorithms are transformed between the GEP and ANN structures during training iterations, and efficient and robust convergence to accurate models is achieved by combining the GEP's global searching and the ANN's gradient optimization capabilities.
In addition, sparsity-enhancing strategies have been introduced to GEPNN to improve the interpretability of the trained models. 
The GEPNN method has been tested for finding different physical laws, showing improved convergence to models with precise coefficients. 
Furthermore, for large-eddy simulation of turbulence, the subgrid-scale stress model trained by GEPNN significantly improves the prediction of turbulence statistics and flow structures over traditional models, showing advantages compared to the existing GEP and ANN methods in both \emph{a priori} and \emph{a posteriori} tests.
\end{abstract}




\end{frontmatter}


\section{\label{sec:introduction}Introduction}
The development of physical models is essential to science and engineering, in which the classical scientific approach is driven by hypotheses. Scientists postulate hypotheses based on observational experience and then revise and improve them by conducting multiple experiments and simulations. 
However, formulating and validating the physical models remains challenging.

With growing numbers of datasets and the development of algorithms and computational resources, data-driven approaches can be applied to physical modelling utilizing raw data from experience or simulations. 
In particular, symbolic regression, which aims to discover explicit expressions through data-driven methods, is suitable for physical model development owing to its strong interpretability. 
Therefore, different machine learning (ML) algorithms have been proposed for symbolic regression tasks, showing progress in various scientific areas.

The concept of sparse regression, which first pre-defines a library of basic functions composing the target model, and then reduces the number of terms in the model expressions and balances model complexity with descriptive ability, is frequently used for symbolic regression tasks. 
For example, sparse identification of nonlinear dynamics (SINDy)~\cite{Brunton2016DiscoveringGE} can discover governing equations in the simplest form under dynamic constraints and balances. 
Moreover, sparse regression of turbulent stress anisotropy (SpaRTA)~\cite{Schmelzer2019DiscoveryOA} has been applied to close the Reynolds-averaged Navier-Stokes (RANS) equations in fluid dynamics with algebraic stress models. 
However, sparse regression must preset a library of candidate functions, which requires sufficient prior knowledge from the users. 

With careful design and a simplified structure, an ANN~\cite{goodfellow2016} can also be used for symbolic regression tasks. The symbolic regression neural network (SRNN)~\cite{Sahoo2018LearningEF, 2020Integration}, which uses fully connected ANN with mathematical operators as activation functions, can develop models with explicit expressions. 
The SRNN can be trained end-to-end through backpropagation and gradient descent. 
However, the fixed network structure, \textit{e.g.} the pre-defined number of hidden layers and mathematical operators, poses limitations on the potential trained expressions, which can significantly affect the performance of the model regression.
AI Feynman~\cite{udrescu2020ai} is a recursive multi-perspective symbolic regression algorithm based on ANN.
By introducing a suite of physics-inspired techniques, such as dimensional analysis and translational symmetry recognition, AI Feynman can partition the symbolic regression problem into simpler sub-tasks and select the relevant input variables for each subtask. Then a brute force algorithm can be appied to try all possible symbolic expression with these variables and give the solution of each subtask. Nevertheless, it cannot create non-trivial constant coefficients, which are essential components of the physical laws. 
Other deep-learning techniques have also been applied to symbolic regression, such as reinforcement learning~\cite{Petersen2021DeepSR}, attentional mechanisms ~\cite{Biggio2021NeuralSR}, \emph{etc.} 
However, the application of deep neural networks is constrained by the characteristics of the ANN structure (\emph{e.g.}, preset network architecture and pre-trained ANN weights), which are difficult to train and inefficient in inference. 

Evolutionary algorithms (EA), including genetic algorithms~\cite{holland1975}, genetic programming~\cite{koza1992, schmidt2009}, \emph{etc.}, apply biology-inspired strategies such as mating, natural selection and evolution, and are probably the most popular tools for symbolic regression tasks. 
Particularly, gene expression programming (GEP)~\cite{ferreira2001} is an advanced type of EA in the sense that a population of individuals is expressed as a nonlinear expression tree, similar to genetic programming, and evolved as a fixed-length chromosome over many generations, similar to genetic algorithm.
Owing to its strong searching ability in the optimization space, GEP has been applied in many areas, such as physical modelling ~\cite{weatheritt2016, li2021data} and engineering ~\cite{Khan2021CompressiveSO, AliKhan2021ApplicationOG}.
Nonetheless, EA-type approaches involve nondirectional optimization because they generate and mutate expressions using biological strategies instead of the information in the training data, such as the data distribution or gradient, which can be extremely inefficient. 
To reduce the dimensions of the search space and further exploit the training data, many researchers have focused on integrating genetic algorithms with other data-driven approaches~\cite{Cranmer2020DiscoveringSM, He2022TaylorGP}. 
Even in these cases, the optimization of constant numbers in expressions also depends on the random combination of symbols or external optimization methods, such as the Broyden-Fletcher-Goldfarb-Shanno (BFGS) method~\cite{1977Quasi}, after training iterations. 
In a recent work, Waschkowski \emph{et al.}~\cite{Waschkowski2022GradientIA} introduced adaptive symbols for model constants in GEP, and the optimization of the constant symbols uses gradient information and thus can be relatively easy to converge. 

To summarize, though different ML methods have been introduced for symbolic regression tasks, there still exist limitations which affect their performance, generalizability and robustness.
In the present study, we propose a new approach called the gene expression programming neural network (GEPNN), with the aim of robust and efficient development of accurate model expressions via symbolic regression.

The key idea for GEPNN is combining GEP and ANNs to compensate for the drawbacks of GEP (\textit{i.e.} slow convergence via stochastic searching of the whole optimization space and difficulties in optimizing constants) and ANNs (\textit{i.e.} dependence on the pre-defined network structures which require sufficient prior knowledge).
We remark that the integration of ANN and EA is not an entirely new idea for the ML community~\cite{Darwish2019ASO, Zhou2021ASO}, and remarkable progress has been shown, especially in areas such as image recognition and natural language processing. 
In particular, a series of methods named as evolutionary neural networks (ENNs) apply EA algorithms to optimize the structures of ANNs, showing the enhanced ability of ENNs for model training.
Nevertheless, to the authors' knowledge, the present study is the first to introduce the idea of combining EA and ANN to develop explicit model expressions via symbolic regression.
This is accomplished by combining GEP with a recently developed type of ANN, the SRNN \cite{Sahoo2018LearningEF, 2020Integration}.  
 
In the GEPNN, the GEP framework generates and searches for candidate expressions using a genetic algorithm, which provides various architectures for SRNN. 
By designing appropriate encoding strategies and transformation principles, the candidate expression can be transformed into a SRNN architecture that can be optimized using gradient information. 
When the SRNN optimization step is completed, the network can be recovered to GEP individuals participating in the evolution of the next generation. 
With the SRNN reinforcement, the optimizing process guided by the local gradient enables the GEPNN to converge more efficiently to models with accurate constants than traditional GEP. 
Meanwhile, genetic operations from GEP can generate and select different expressions to build SRNN, which makes up for the disadvantage that SRNN needs to preset the network structure using numerous trial-and-error experiments. 
This novel method will be extensively tested in a series of cases, demonstrating the advantages of GEPNN compared to traditional GEP and SRNN methods in developing physical models with explicit expressions.

The outline of this paper is as follows. First, we introduce GEP and SRNN in Sec.~\ref{sec:method}, which are the fundamental methodologies used in this study. Subsequently, the novel GEPNN method is proposed in Sec.~\ref{sec:GEPNN}, which combines the advantages of EA and ANN. Furthermore, GEPNN is applied to determine the algebraic equations of the physical models described in Sec.~\ref{sec:results}, including three physical equations and a closure modelling problem in the large-eddy simulation of turbulence. Finally, we conclude this paper in Sec.~\ref{sec:conc}.

\section{\label{sec:method}Fundamental Methodology}
In this section, we briefly introduce GEP and SRNN, the basic algorithms used in this research. Then, we compare their performance using a polynomial case and discuss the limitations of both methods, motivating the introduction of the novel method GEPNN in the present study.

\subsection{\label{subsec:GEP}Gene expression programming}
GEP~\cite{ferreira2001} is a powerful EA which can solve a series of problems, such as symbolic regression, sequence induction with and without constant creation, block stacking, \emph{etc.} For the symbolic regression task, a population of mathematical expressions is evolved through natural selection and genetic operations over many generations. Unlike traditional genetic algorithms, the fundamental advantage of GEP lies in its unique design of the individual, \emph{i.e.}, the chromosome, which contains mathematical expression information. Each chromosome has a distinct genotype and phenotype. The purpose of the genotype is to store genetic information that can be evolved in each generation, while the phenotype corresponds to the explicit expression extracted from the genetic information.

The genotype of the chromosome is encoded as fixed-length strings, which consist of one or more genes and a set of linking symbols to link genes. A single gene can represent a mathematical expression with a string of symbols. For example, a polynomial of $x$ and $y$
\begin{equation}
\label{eq:gene_equation}
(x+1)\times(y-1)
\end{equation}
can be encoded as a single gene with a genotype string
\begin{equation}
\label{eq:gene_genotype}
<\times ~+ ~- ~x~1~y~2>,
\end{equation}
where the mathematical symbols $\times$, $+$, and $-$ represent multiplication, addition, and subtraction, respectively, the alphabetic characters $x$ and $y$ represent the variables, and $1$ and $2$ represent constant numbers. 

\begin{figure}[!ht]
\centering
\includegraphics[width=6cm]{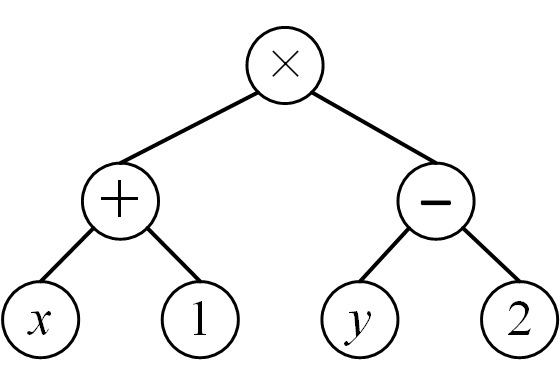}
\caption{Example ET corresponding to Eqn.~(\ref{eq:gene_equation})}
\label{fig:gene_phenotype}
\end{figure}

The genotype string in GEP can be decoded into its phenotype, \textit{i.e.} a mathematical expression, with an expression tree (ET) structure. 
As shown in Fig.~\ref{fig:gene_phenotype}, the genotype string given by Eqn.~(\ref{eq:gene_genotype}) can be decoded by taking the root symbol, filling its child nodes with the following $n$ symbols, and repeating until no more symbols can be added to this tree. 
Here, $n$ represents the arity of the symbol, \emph{e.g.}, multiplication $\times$ takes two arguments, so $n_{\times}=2$.

\begin{figure}[!ht]
\centering
\includegraphics[width=4cm]{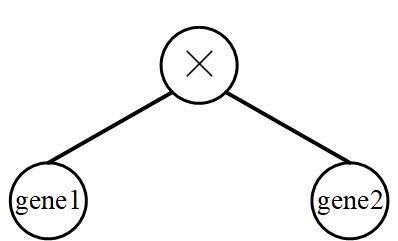}
\caption{Example ET corresponding to the chromosome with multiple genes}
\label{fig:chromo_phenotype}
\end{figure}

By combining multiple genes, the chromosome can represent long and complex expressions. It can obtain its phenotype by decoding its genes separately and connecting those ETs with link symbols. Consider a chromosome with two genes $<gene1>$, $<gene2>$ and a link symbol $\times$. The genotype of this chromosome can be written as
\begin{equation}
\label{eq:chromo_genotype}
\times~<gene1>~<gene2>.
\end{equation}
The phenotype can be expressed as shown in Fig.~\ref{fig:chromo_phenotype}. An expression described by the ET can be calculated directly from the leaf nodes to the root node, which is easy to implement in the program. 

Due to the design of the genotype, a chromosome can easily participate in genetic operations because all the genetic operations applied to a chromosome are equivalent to replacing part of the gene with another fixed-length string. When this chromosome needs to be evaluated, it can be recursively decoded to the phenotype and efficiently calculated as a regular mathematical expression.

Compared to other data-driven algorithms, for example, ANN~\cite{goodfellow2016, 2018Physics} and random forest~\cite{2004Machine}, the advantage of GEP is its global searching strategy. GEP can generate complex mathematical expressions by applying simple genetic operations, and the variation in chromosomes allows exploring the entire search space. Thus, GEP is a global optimization algorithm that can search the optimization space and has the potential to address the problem of function discovery and symbolic regression. However, genetic operations do not consider the optimization direction from the training data, leading to convergence difficulties. Moreover, constants in the expression depend on the combination of preset symbols, making it difficult to obtain accurate values of the constant numbers~\cite{Ryan2003AnAO, Zhong2017GeneEP}.

\subsection{\label{subsec:SRNN}Symbolic regression neural network}
\begin{figure}[!ht]
\centering
\includegraphics[width=13cm]{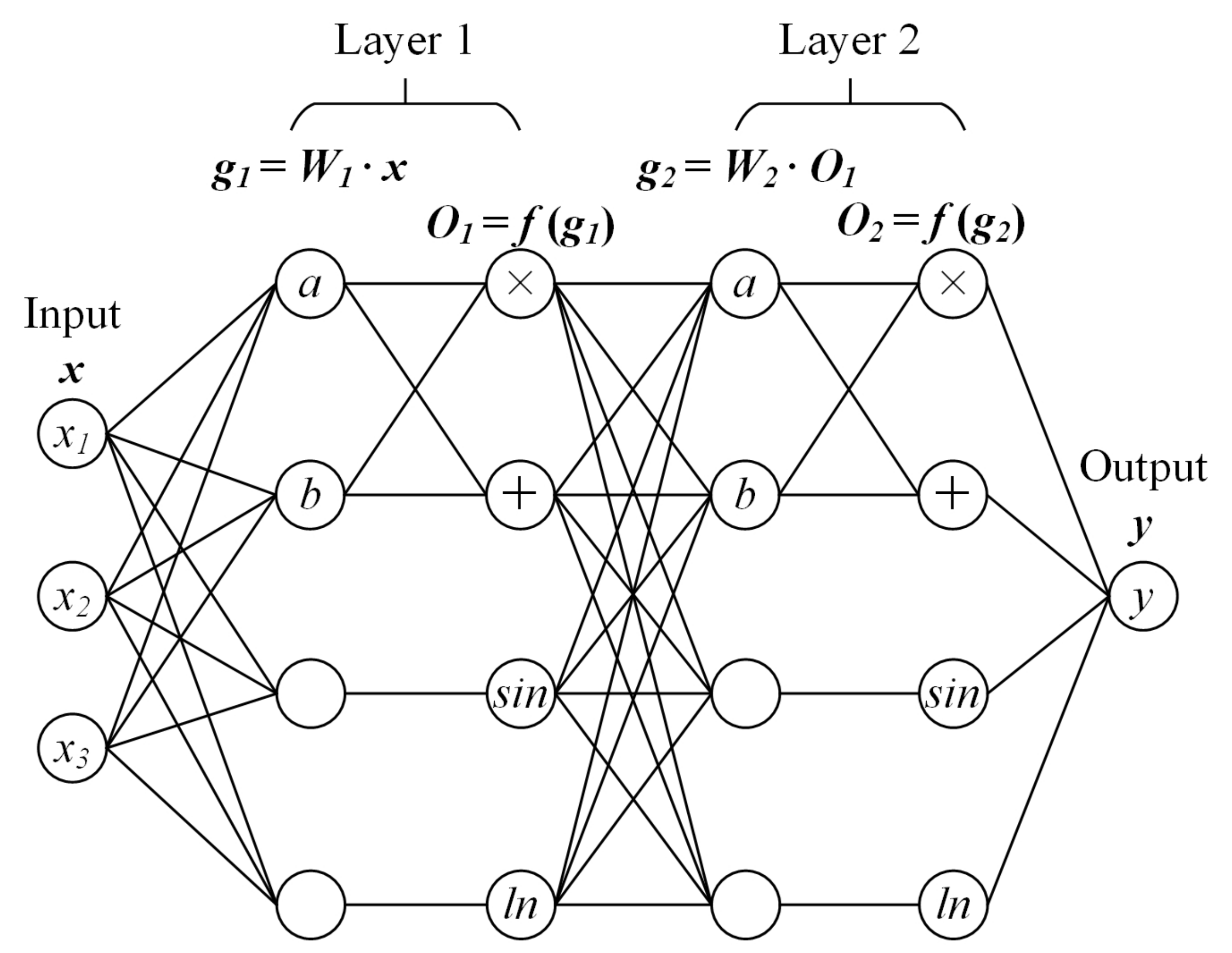}
\caption{Example of an SRNN with two hidden layers and four activation functions.}
\label{fig:symreg}
\end{figure}

SRNN~\cite{2020Integration} is an ANN architecture recently designed for symbolic regression tasks. 
Fig.~\ref{fig:symreg} exhibits the schematic of an example SRNN, with two hidden layers for visual simplicity.
The basic structure of SRNN is usually a fully connected ANN, in which the $i^{th}$ layer of the network can be derived from the $(i-1)^{th}$ layer:
\begin{equation}
\begin{aligned}
&g_{i}=W_{i}I_{i-1},\\
&O_{i}=f(g_{i}),
\end{aligned}
\end{equation}
where $I_{i-1}$ is the input data from the $(i-1)^{th}$ layer, $W_{i}$ is the weight matrix of this connection, and $O_{i}$ is the output data of the $i^{th}$ layer.
It is noted that in SRNN, the activation functions $f(*)$, which are typically nonlinear mappings like sigmoid or Relu~\cite{2011Deep} in traditional ANNs, are simple mathematical operators.
Specifically, the activation function $f(*)$ in the layer $i$ may consist of separate functions for each node of layer $i-1$ (such as $sin$ or $ln$) and may include functions that can adopt more than one argument to generate one output (such as the multiplication function). 

SRNN can fit complex combinations and compositions of the input variables with various primitive functions by stacking multiple layers and applying different activation functions. It can be trained end-to-end with backpropagation and other powerful deep learning techniques while still producing interpretable expressions. However, gradient-based methods suffer from a gradient-loss problem at the local optimum or saddle point during training. 
Moreover, the network structure, which can significantly affect the performance of the training, needs to be decided before training and is not flexible, thus limiting the application of this approach~\cite{Zhan2022, Kronberger2021ShapeConstrainedSR}.

\subsection{\label{subsec:GEPvsSRNN}A polynomial test case for GEP and SRNN}
In order to show the advantage and disadvantage of the GEP and SRNN methods, we test their performance for a simple polynomial case.
Consider the following polynomial expression:
\begin{equation}
\label{eq:polynomial}
y=((((x_{1}*C_{1})*(x_{2}+C_{2}))*(((C_{3}+x_{1})+(x_{2}*C_{4})))-(C_{5})),
\end{equation}
where $x_{1}$ and $x_{2}$ are the input variables, and $C_{1,\dots, 5}$ are constants. 
By setting the constants as 
\begin{equation}
\label{eq:poly_consts}
C_{1}=0.4, C_{2}=1.25, C_{3}=-0.85, C_{4}=1.25, C_{5}=1.7,
\end{equation}
Eqn.~(\ref{eq:polynomial}) can be expanded and simplified as follows:
\begin{equation}
\label{eq:poly_expand}
y=0.4x_{1}^{2}x_{2} + 0.5x_{1}^{2} + 0.5x_{1}x_{2}^{2} + 0.285x_{1}x_{2} - 0.425x_{1} - 1.7,
\end{equation}
which is used to generate a training dataset with $x_{1}, x_{2}\sim N(0,1)$. 
Note that random noise with a magnitude of $10\%$ is then superposed to the outputs y.

For GEP, the chromosome of the expression in Eqn.~(\ref{eq:polynomial}) contains five genes, and the genotype can be written as
\begin{equation}
\begin{aligned}
\label{eq:poly_gene}
&-~\times~+~\times, \\
<\times~x_{1}~C_{1}><+~x_{2}~C_{2}>&<+~C_{3}~x_{1}><\times~x_{2}~C_{4}><C_{5}>.
\end{aligned}
\end{equation}
The initial population for the GEP training is randomly generated, except for that an individual with chromosome Eqn.~(\ref{eq:poly_gene}) is created and inserted into the initial population, which ensures that the starting point of GEP and SRNN is consistent. 
The values of $C_{1,\dots, 5}$ are initialized with a normal distribution $N(0,1)$.
It is noteworthy that multiple genotypes can exist for the same expression. 

\begin{figure}[!ht]
\centering
\includegraphics[width=14cm]{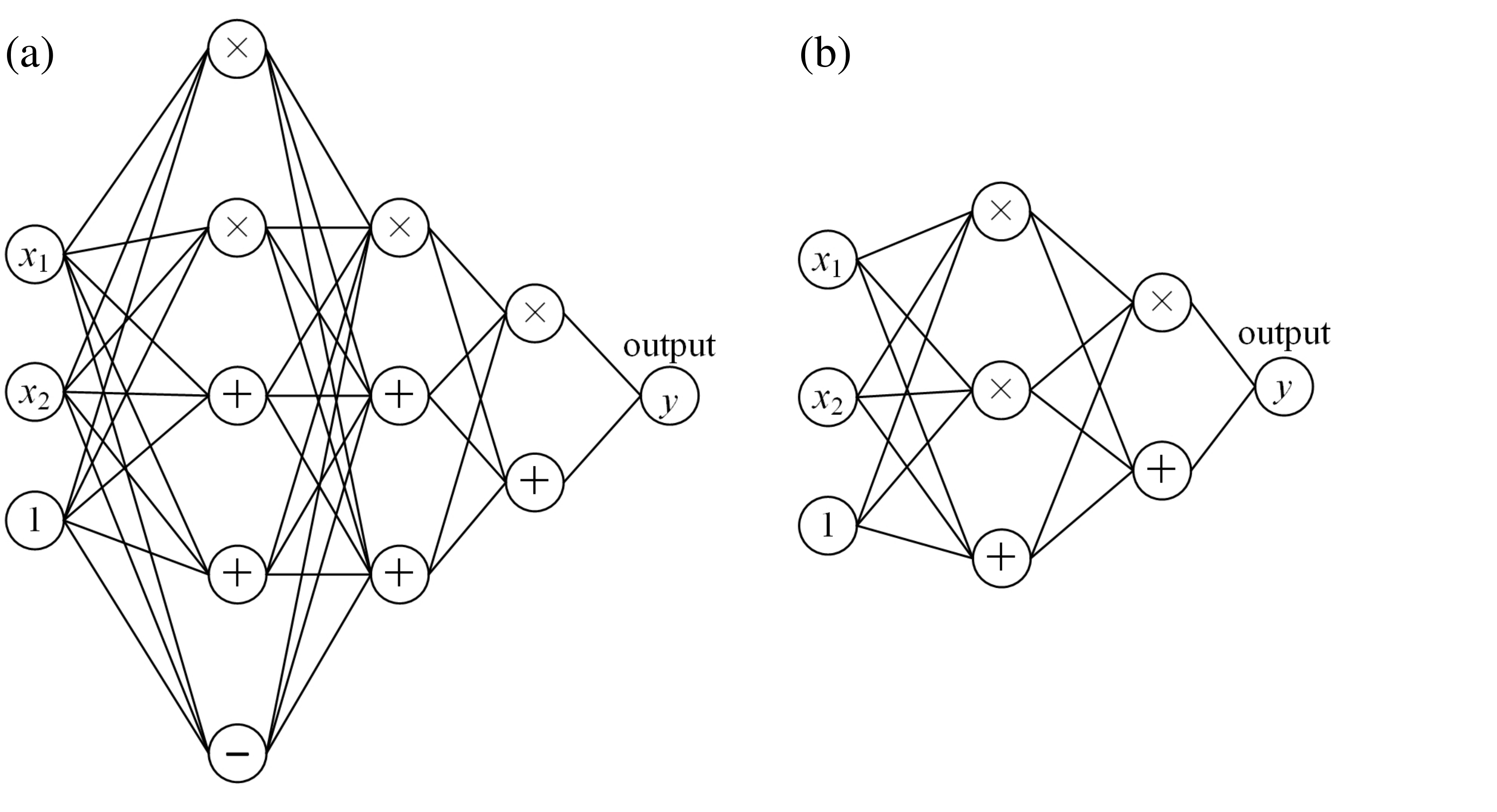}
\caption{SRNN structures used for the polynomial test case: (a) SRNN1, a specially designed structure that can represent the polynomial case; and (b) SRNN2, a simpler structure with weak exploring ability.}
\label{fig:srnn_simpler}
\end{figure}

Furthermore, two different SRNNs are built to reflect the influence of the ANN architecture on the performance, as presented in Fig.~\ref{fig:srnn_simpler}. 
SRNN1 is designed explicitly with the prior knowledge of Eqn.~(\ref{eq:polynomial}), and the mathematical operators are selected to ensure that SRNN1 can represent this polynomial. 
Thus, SRNN1 can easily represent Eqn.~(\ref{eq:polynomial}) by setting the special connection weights. 
By contrast, SRNN2 is much simpler and arbitrary, leading to a limited exploration space. The connection weights of SRNN1 and SRNN2 are also initialized with a normal distribution.

\begin{figure}[!ht]
\centering
\includegraphics[width=10cm]{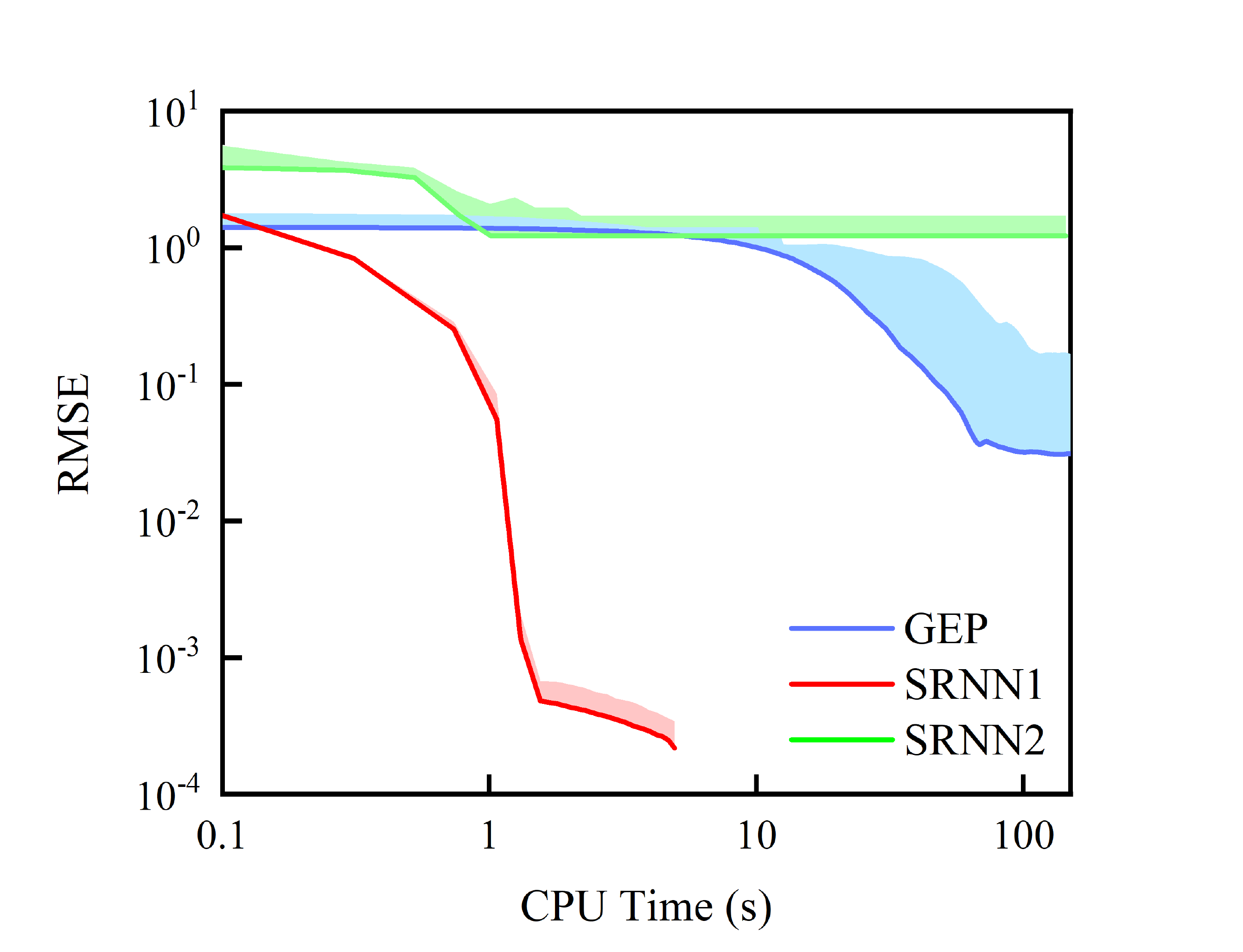}
\caption{Convergence comparison for the polynomial test case. The Y-axis is the root-mean-squared error (RMSE) between the model output $y^{*}$ and data $y$. The blue, green, and red areas show the results from ten training cycles for each method, and the solid lines represent the best result for each method.}
\label{fig:genn_poly}
\end{figure}

Fig.~\ref{fig:genn_poly} compares the training processes of the GEP, SRNN1, and SRNN2 for Eqn.~(\ref{eq:polynomial}). The root-mean-squared error (RMSE) of SRNN1 decreases quickly during the training stage, and its training terminates after $1,000$ epochs, reaching a low error value. 
Owing to its simpler architecture, SRNN2 cannot identify the correct form of the target expression, leading to convergence to a higher RMSE, even after $30,000$ epochs of training.
On the other hand, SRNN1 can extract constants from the corresponding connection weights
\begin{equation}
\label{eq:poly_genn}
C_{1}=0.398, C_{2}=1.247, C_{3}=-0.852, C_{4}=1.243, C_{5}=1.766,
\end{equation}
which are very close to the target values of Eqn.~(\ref{eq:poly_consts}). 
This shows that using the gradient information of the training data, the SRNN can determine the optimization direction and adjust the constants accordingly. 
By setting a reasonable network structure, SRNN1 converges to the best result. 
However, with an improper SRNN structure, the SRNN2 shows much worse results.

As a comparison, the RMSE value given by the GEP training results in Fig.~\ref{fig:genn_poly} is still quite high after $1,000$ generations, which costs much more computational resources compared to SRNN1. The expanded form of the GEP result is
\begin{equation}
\label{eq:poly_gep}
y=0.430x_{1}^{2}x_{2} + 0.443x_{1}^{2} + 0.430x_{1}x_{2}^{2} + 0.443x_{1}x_{2} - 0.341x_{1} - 1.651,
\end{equation}
which represents the same functional form as in Eqn.~(\ref{eq:poly_expand}). However, the constant values of the GEP expressions are inaccurate because searching the global optimization space and generating precise constant numbers by traditional genetic operations are inefficient. 

The results of the polynomial case reflect two aspects of the symbolic regression task: one is to uncover the underlying expression structure, which is an advantage of GEP; and the other is to generate a proper set of constant coefficients in the expression, which can be accomplished efficiently and accurately using the SRNN. 
At the same time, the presented results also suggest that the GEP cannot consider the optimization direction, leading to difficulty in obtaining accurate constants and thus slowing the overall convergence rate. 
The SRNN performance, on the other hand, is greatly affected by the preset network structure because the structure defines the search space and the most complex expression form that the SRNN can discover. However, there is usually a trade-off between model complexity and network performance, and the setting of the network structure usually requires prior knowledge of the application fields.
Therefore, we intend to develop a novel method for general symbolic regression tasks in the present study by combining the advantages of GEP and SRNN, with the capability of efficiently discovering the expression structure and regressing the model coefficients at the same time.

\section{\label{sec:GEPNN}Gene Expressions Programming Neural Network}
The GEPNN method introduced in the present study combines the GEP framework with the optimization step of the SRNN. 
To be specific, the GEP framework generates expressions using an evolutionary algorithm, and the SRNNs are built with the expressions generated by GEP and responsible for efficiently optimizing the expressions with gradient information. 
The key component of this method is the transformation of expressions between the GEP algorithm and the SRNN, which contains a special design for GEP individuals and the transformation principles. 
Based on that, the GEPNN framework, in which the SRNN optimization is incorporated in the GEP iterations, is introduced.
In addition, we propose sparsity techniques applied to GEPNN, which reduce the complexity of trained expressions.
Finally, we cover the implementation of GEPNN and the selection of hyperparameters.

\subsection{\label{subsec:Trans}Transforming expressions between GEP and SRNN}
The critical challenge for combining GEP with SRNN is to transform expressions between GEP and SRNN.
An expression in GEP is encoded as the genotype of a chromosome, which is essentially an ordered string. However, in a SRNN, a structured network with weighted connections contains all the information of the expressions. To optimize the expression with the SRNN, we need to transform the genotype of chromosomes into ANN architectures. In other words, the SRNN should be initialized using GEP chromosomes. 
At the same time, to continue the GEP training iterations after the SRNN optimization step, we need to be able to transform the SRNN back into the GEP genotype without breaking the original structure of the chromosome. 
Therefore, the transformation of expressions between GEP and SRNN requires a special design for GEP individuals and the transformation principles.

\begin{figure}[!ht]
\centering
\includegraphics[width=13cm]{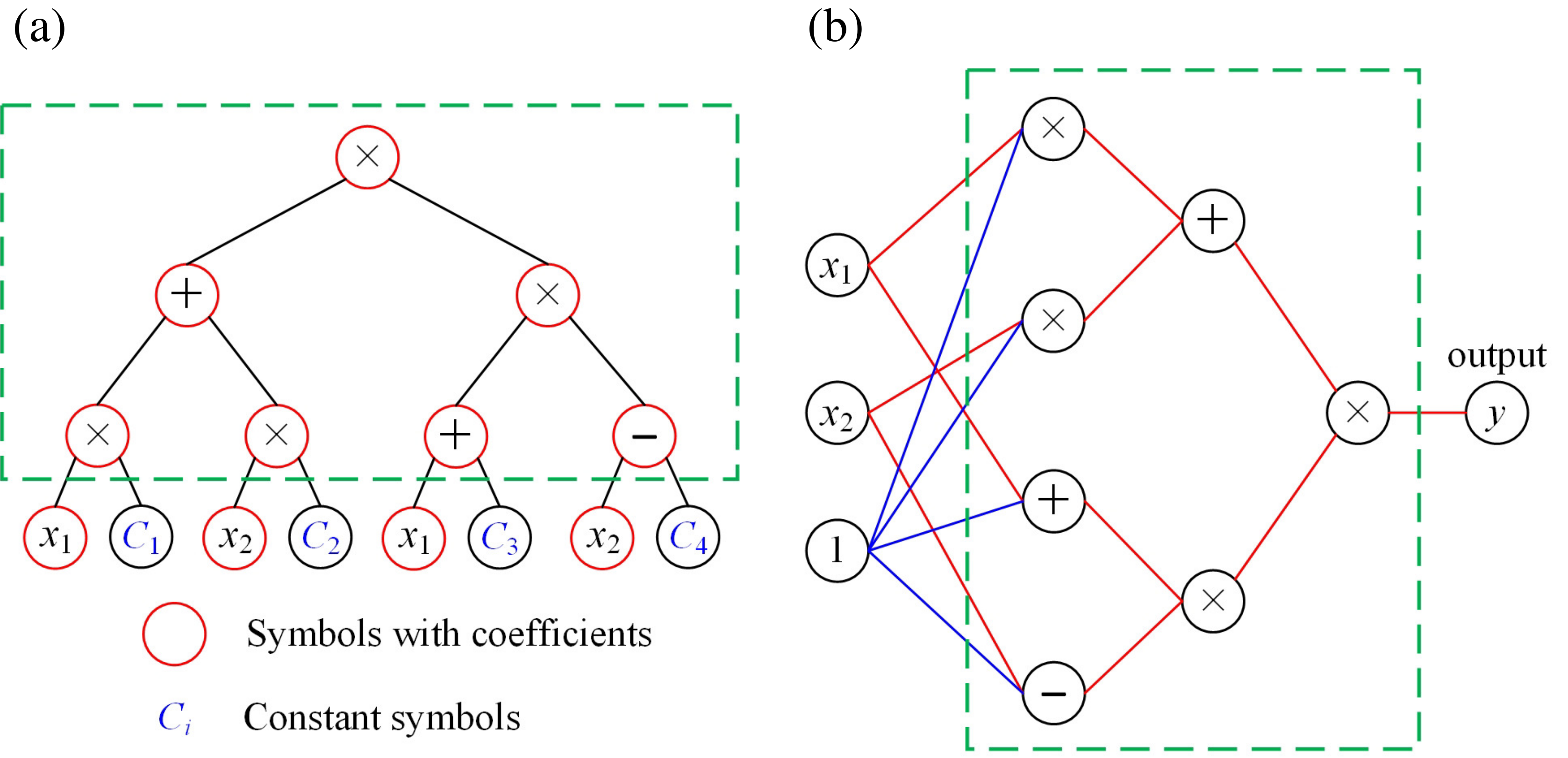}
\caption{ET and corresponding architecture of the SRNN of Eqn.~(\ref{eq:genn}). (a) ET of the chromosome; and (b) architecture of the SRNN.}
\label{fig:genn_arch}
\end{figure}

Fig.~\ref{fig:genn_arch} shows the design of GEP and SRNN architectures for transformation.
Essentially, the transformation of the GEP into the SRNN is a process similar to decoding a chromosome into its ET, which requires a recursive analysis of each symbol. 
As an example, the expression
\begin{equation}
\label{eq:genn}
[(x_{1}\times C_{1})+ (x_{2}\times C_{2})]\times [(x_{1}+ C_{3})\times (x_{2}- C_{4})]
\end{equation}
has the following genotype:
\begin{equation}
\label{eq:genn_genotype}
\times~<+~\times~\times~x_{1}~C_{1}~x_{2}~C_{2}>~<\times~+~-~x_{1}~C_{3}~x_{2}~C_{4}>,
\end{equation}
where $<*>$ represents a gene, $C_{i}$ are constant numbers, and $x_{i}$ denote the input variables.
As discussed in Section~\ref{subsec:GEP}, the genotype in Eqn.~(\ref{eq:genn_genotype}) can be decoded into the phenotype in Eqn.~(\ref{eq:genn}) via the ET shown in Fig.~\ref{fig:genn_arch}(a), which is then directly linked to the SRNN architecture with input neurons $x_{1}$, $x_{2}$ and $1$ shown in Fig.~\ref{fig:genn_arch}(b).

Specifically, the variable symbols $x_i$ in chromosomes are transformed into the input neurons in SRNN, and the operator symbols $\{+,-,\times\}$ are converted as hidden neurons and connected according to the sequence of chromosomes. 
Moreover, the constant symbols $C_i$ in GEP chromosomes are transformed into the connections between the input neuron $1$ and the hidden layers, indicated by the blue solid lines in Fig.~\ref{fig:genn_arch}~(b). 
In addition to the existing constant symbols in the GEP chromosome, each connection in SRNN, which is indicated by the red solid lines in Fig.~\ref{fig:genn_arch}~(b), contains a trainable weight. Thus, we can determine a special coefficient for each symbol, shown as red nodes in Fig.~\ref{fig:genn_arch}~(a).
These coefficients are initialized with a value $1$ and can be optimized by SRNN. 

Based on the principle listed above, the expressions can be converted from GEP to SRNN as shown by the corresponding ET and SRNN structures in Fig.~\ref{fig:genn_arch}.
Thereafter, the resulting SRNN can be trained end-to-end using the gradient information obtained from the training data to optimize the trainable weights. 
Benefitting from the transformation of the GEP into the SRNN, the constant values in the mathematical expression of the chromosome can be promptly optimized to their local optimum.
It should be noted that as traditional GEP has difficulty training constant coefficients. The number of constants in GEP models is usually limited by the preset constant symbols and the maximum length of chromosomes, and even inaccurate unless with exceptionally long training, as discussed in Section~\ref{subsec:GEPvsSRNN}.
Therefore, the introduction of coefficients for every symbol is usually not considered in traditional GEP. 
The extension of the symbol coefficients brings a stronger exploration ability for GEPNN, which is expected to significantly enhance the expressivity of chromosomes, especially the ability to train accurate model coefficients. 

It is also noted that the structure of the hidden layers of the SRNN is identical to the green-boxed part of the ET in Fig.~\ref{fig:genn_arch}(a).
Therefore, after the SRNN optimization, the transformation of the SRNN back to the GEP ET and hence the chromosome is straightforward.
Subsequently, the expression optimized by SRNN can be included in the evolution of the next GEP generation, and new created expression structures generated by GEP can be further optimized based on the GEP-SRNN transformation in this iterative process.  
In this way, by combining GEP with SRNN, we aim to efficiently and accurately train the expression structure and the coefficients at the same time. 

\subsection{\label{subsec:GEPNN}The GEPNN framework}
\begin{figure}[!ht]
\centering
\includegraphics[width=12.5cm]{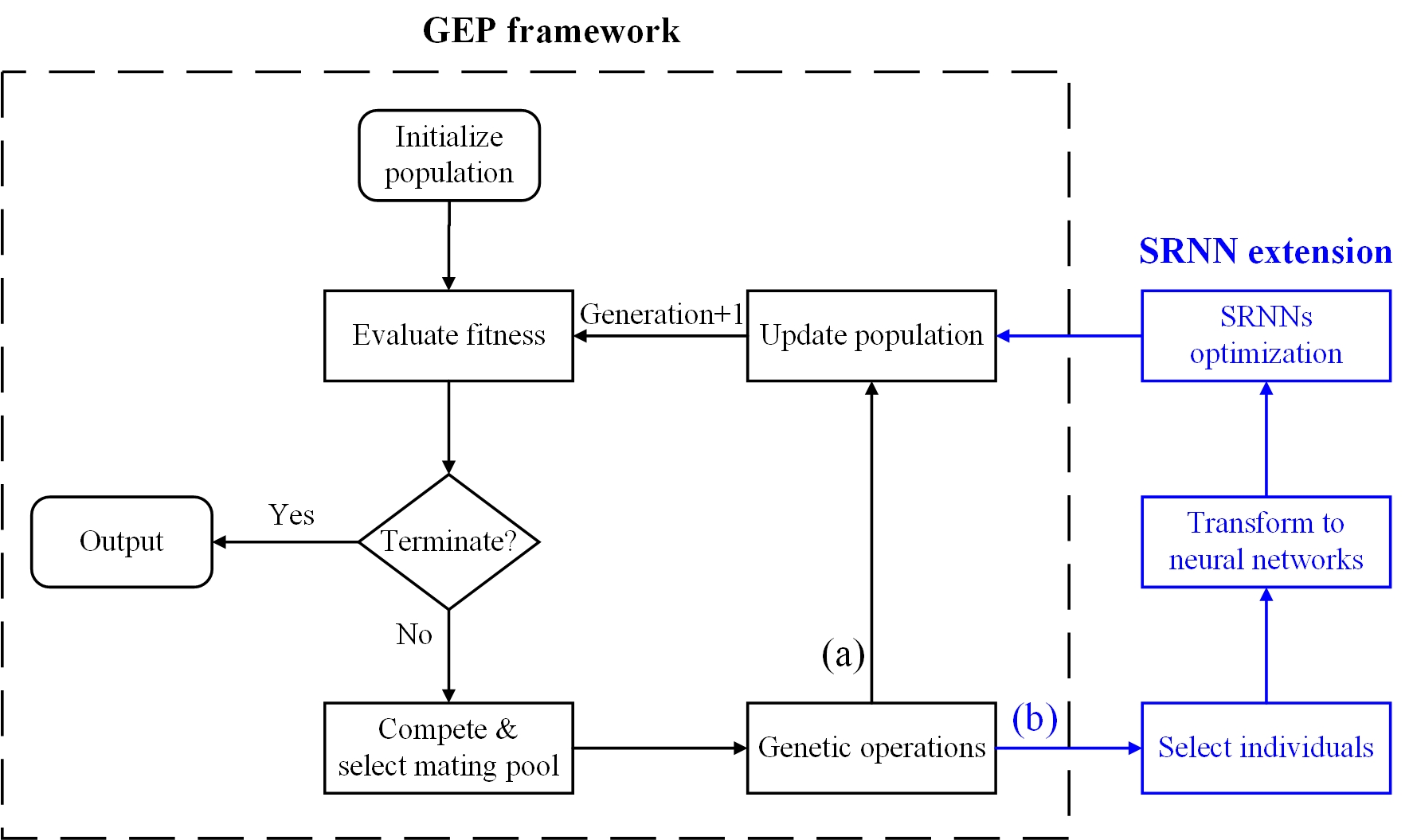}
\caption{Flowchart of the GEPNN framework. (a) Traditional GEP framework, and (b) the optimization steps with SRNN.}
\label{fig:gepnn_flowchart}
\end{figure}

Based on the transformation process of expressions between GEP and SRNN discussed above, the framework of GEPNN is introduced in Fig.~\ref{fig:gepnn_flowchart}, in which the expression structures are determined using genetic algorithms, and the constants are efficiently optimized with SRNN.
Initially, a population of individuals is randomly created. Each individual’s fitness is then evaluated according to the user-defined loss function, \textit{i.e.} the training objective. 
If the termination criterion (\emph{e.g.}, achieving a preset generation number or obtaining a loss lower than a preset value) is not fulfilled, individuals compete with each other depending on their fitness, and the winner is selected for the mating pool individuals. 
Next, the individuals are updated by applying genetic operations to the mating pool. 
For the traditional GEP method, the updated population is evaluated, and the training process continues over the next generations. 
However, for the GEPNN, with the integration of the SRNN, the individuals are selected, transformed and optimized using ANNs. 
After SRNN optimization, these ANNs are recovered to chromosomes and updated the corresponding individuals to participate in the subsequent generations.

Traditional genetic operations in GEP include mutation and crossover. 
All individuals in the mating pool can perform one or more operations to manipulate and change their chromosomes. 
We remark that the SRNN optimization introduced in the GEPNN can be considered as an additional genetic operation specifically designed to enable efficient optimization of the constants in GEP chromosomes.
Therefore, in the GEP iterations, all individuals in the mating pool have the chance to perform the genetic operation and SRNN optimization.
In addition, $m$ candidates with the highest fitness values, i.e., those with the best expressions, are automatically subjected to SRNN optimization. Thus $m$ is a hyperparameter that depends on the population of the GEPNN and computational resources.

It is noteworthy that SRNN optimization is not applied in every generation of the GEPNN, as it incurs additional computational costs and is usually unnecessary. 
The SRNN optimization is applied for every $n$ generations, and $n$ is another hyperparameter that can be modified depending on the training situation, for example, training data size, and complexity of the training target. 
In addition, the SRNN optimization is applied when a new expression structure with better fitness values appears after other genetic operations. 

\begin{figure}[!ht]
\centering
\includegraphics[width=12cm]{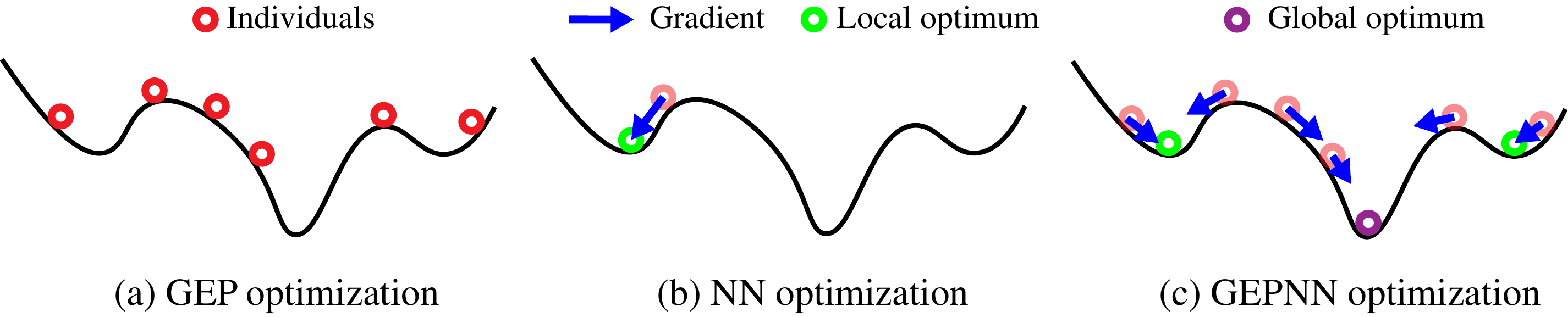}
\caption{Optimization process of different algorithms. (a) Global searching with GEP, (b) local optimization with ANN, and (c) global optimization with local optimization abilities using GEPNN}
\label{fig:gepnn_grad}
\end{figure}

In principle, GEPNN compensates for the disadvantages of GEP and SRNN. 
Fig.~\ref{fig:gepnn_grad} shows a schematic of the optimization processes of the GEP, SRNN, and GEPNN. 
GEP uses genetic operations for mathematical expressions to create new forms of equations for global searching. 
However, there is no direction for optimization in the subsequent generation. 
Unlike GEP, SRNN uses a local gradient to instruct the equations to descend to the local optimum. 
However, it may fall into a local optimum or saddle point without gradients, and the network structure is critical to its performance as shown in Section~\ref{subsec:GEPvsSRNN}. 
By integrating GEP with the SRNN, GEPNN combines the advantages of the two methods to obtain diverse individuals that can be optimized with gradients, finally finding the global optimum efficiently and accurately. 

\subsection{\label{subsec:sparsity}Sparsity and non-coding symbol}
In addition to the GEPNN framework, various strategies have been introduced to enhance the sparsity of the trained models.
We remark that enhancing model sparsity is usually desired for ML tasks, as it has the potential to make the trained model more interpretable and also avoid overfitting ~\cite{Waschkowski2022GradientIA}.
For the GEPNN framework, ANNs are incorporated in the training iteration, so it is natural to apply existing strategies for the sparsity of ANNs here.
The details of the sparsity strategies introduced to the present GEPNN method are thus discussed below.

First of all, for the training of SRNN, a regularization term is introduced to the loss function, and the intention is to set as many connection weights to $0$ as possible so that those connections can be removed from the final trained model expression.
Adding a regularization term to the loss function is a straightforward and popular method to enforce the sparsity of neural networks \cite{Xu2010L12R, Fan2014ConvergenceOO}. 
The commonly used $L_{q}$ regularization acts as:
\begin{equation}
\label{eq:lqr}
L_{q}(W) = \sum_i \lvert w_i \rvert^{q},
\end{equation}
where $w_i$ denotes the connection weights in the neural network. 
In the present study, we use a smoothed $L_{0.5}$ regularization \cite{2020Integration} to avoid the singularity of the gradient as the weights go to $0$ 
\begin{equation}
L_{0.5}(w)=
\left\{
\begin{array}{lr}
\lvert w \rvert^{1/2} & \lvert w \rvert \geq a,\\
(-\frac{w^{4}}{8a^{3}}+\frac{3w^{2}}{4a}+\frac{3a}{8})^{1/2} & \lvert w \rvert \le a.
\end{array}
\right.
\end{equation}
Here, $a$ is the transition point between the standard $L_{0.5}$ regularization and the smoothed function, which is set to $a=0.01$ in our experiments.

Second, we use a multiphase training strategy following \cite{2020Integration} and \cite{Martius2016ExtrapolationAL} for the SRNN optimization, which divides the training into two stages. In the first stage, a standard learning rate is set to optimize all the connection weights in the SRNN. After this stage, weights below a certain threshold $\beta$ are set and fixed as $0$, which filters the small-magnitude weights out and enforces the sparsity of SRNN. Then, in the second stage, the training of the filtered SRNN continues with a reduced learning rate to fine-tune the remaining weights. This multiphase training strategy ensures the sparsity of SRNN and allows the sparse network to be fully trained.

\begin{figure}[!ht]
\centering
\includegraphics[width=12.3cm]{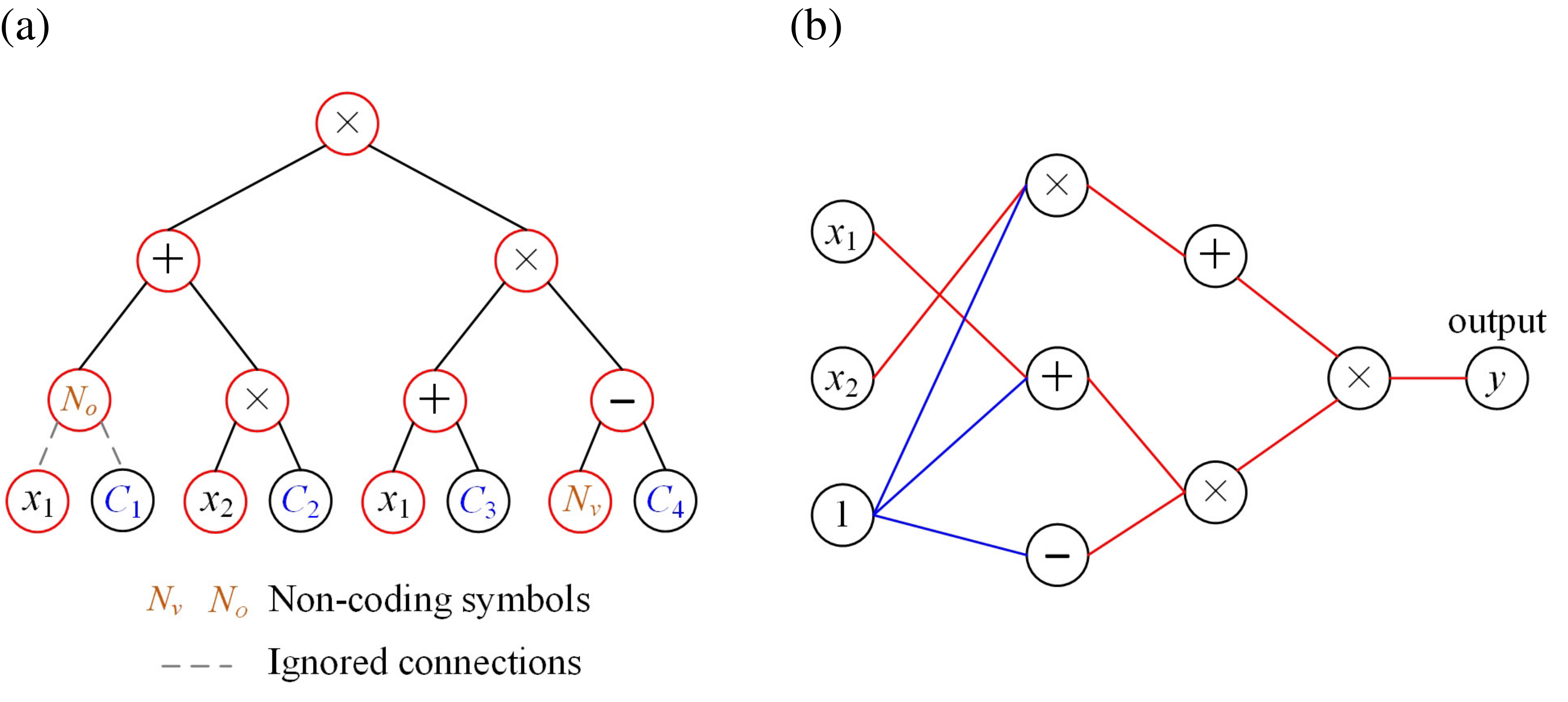}
\caption{ET with pass symbols and corresponding architecture of the SRNN. (a) ET of the chromosome; and (b) pruned architecture of the SRNN.}
\label{fig:prun_arch}
\end{figure}

Finally, it is important to make sure that the sparsity enforced during the SRNN optimization can be kept after the expressions are transformed into GEP chromosomes.
Therefore, we introduce non-coding symbols to prune the chromosome. 
The non-coding symbol is defined here as a symbol without expressive function.
Similar to the GEP algorithm itself, the idea of non-coding symbols is also from the biology field, in which the non-coding DNA represents the sequences that do not encode proteins. 
A non-coding symbol in GEPNN will be created only during the expression transformation from SRNN to GEP. 
In this process, if a connection weight in SRNN is below a threshold $\beta$, its corresponding symbol in the chromosome will be replaced with a non-coding symbol, which means the coefficient of this symbol is small enough to be ignored. 
In particular, to maintain the structure of the chromosomes, the non-coding symbols are distinguished into two categories, \textit{i.e.} the non-coding operators $N_{o}$ which have more than one arity and will take the place of the operator symbols, and the non-coding variables $N_{v}$ which will only appear on the leaf nodes. 
Fig~.\ref{fig:prun_arch} shows an example ET and the corresponding SRNN structure of a pruned expression. 
While calculating its fitness, the non-coding symbols return value $0$, which act as pruning. 
By introducing non-coding symbols, the sparsity of GEPNN is further strengthened. 
Note that the non-coding symbols can be replaced with valid symbols via genetic operations in GEP, which ensures that the pruned expression has the potential to evolve in the following generations.

With the regularization and non-coding symbols, GEPNN can obtain simpler and more interpretable expressions without affecting the training of SRNN and the evolution of GEP.

\subsection{\label{subsec:implementation}Implementation and hyperparameters}
The GEP framework employed in this study was developed by Weatheritt and Sandberg~\cite{weatheritt2016}, and its implementation was programmed using Python~\cite{van1995python}. The SRNN used in this study is implemented via the Tensorflow, which is a powerful framework for deep learning. In the following paragraphs, we will illustrate the details and hyperparameters of GEPNN.

The loss function of the SRNN is defined using the mean-squared error between the prediction $\hat{y}$ and label $y$ using the mini-batch gradient descent algorithm, which can be written as
\begin{equation}
Loss=\frac{1}{N} \sum_{i}^{N} (\hat{y}_{i}-y_{i})^{2},
\end{equation}
where $N$ denotes the batch size. The Adam optimizer~\cite{Kingma2015AdamAM}, which considers the historical effects of gradients, is used to update the weights in the ANN. 

For gradient-based approaches, the learning rate is an important hyperparameter. In the early training process of GEPNN, the learning rate of SRNN is set to a relatively large value to accelerate the optimization, while for later stages the learning rate of SRNN needs to be fine-tuned. 
Thus, the learning rate changes during the GEP training as $L_{r}^{(g)}={L_{r}}/{g}$, where $g$ refers to the $g^{th}$ generation of GEP iterations and $L_{r}$ is the global learning rate as a hyperparameter.
Furthermore, for each GEP generation with fixed $g$, the learning rate of each epoch in the SRNN optimization exponentially decays as follows:
\begin{equation}
\label{eq:decay_lr}
L_{r}^{*}=L_{r}^{(g)}\times(d_{r})^{{g_{e}}/{d_{e}}},
\end{equation} 
where $L_{r}^{*}$ and $g_{e}$ are the learning rate and global epoch number in the current training step, respectively. 
Moreover, $L_{r}^{(g)}$ is the initial learning rate in this generation, and $d_{r}$ and $d_{e}$ are hyperparameters representing the decaying rate and decaying epoch, respectively. 
It is noted that with the decaying learning rate, the training speed at the beginning is relatively fast, while numerical stability can be ensured approaching the final result.

The number of training epochs of each SRNN optimization step are adjusted dynamically according to convergence. After a fixed number of training epochs, we use the loss value to determine whether the training is converged. The training process of SRNN will be terminated if (1) the current loss is less than a preset threshold, (2) the loss value no longer decreases, or (3) the training epochs achieve the preset maximum epochs. 

Moreover, the threshold $\beta$ used for masking the SRNN and pruning the expression in Section~\ref{subsec:sparsity} is dynamically adjusted according to the magnitude of weights, defined as follows:
\begin{equation}
\label{eq:threshold}
\beta = b \sum_i \lvert w_{i} \rvert,
\end{equation}
where $b$ is a hyperparameter denoting the threshold scale.

The settings of all hyperparameters are listed in Table~\ref{tab:GEPNN_setting}. We set a large value for total generations to ensure that the GEP and GEPNN converges during the training stage. Owing to the different training expenses of genetic algorithms and ANNs, we use the CPU time consumed during training to compare the performance of each method. 

\begin{table}[!htb]
\centering\small
\caption{\label{tab:GEPNN_setting}Hyperparameters for GEP and SRNN in experiments}
\begin{tabular}[b]{ll}
\specialrule{0.05em}{3pt}{3pt}
GEP & Value \\\specialrule{0.03em}{3pt}{3pt}
Maximum generations & $1,000$ \\
Population size & $400$ \\
Mating pool size & $100$ \\
Tournament size & $2$ \\\specialrule{0.03em}{3pt}{3pt}
SRNN & Value \\\specialrule{0.03em}{3pt}{3pt}
Batch size $N$ & $5,000$ \\
Fixed epochs & $5,000$ \\
Maximum epochs & $10,000$\\
Loss value threshold & $10^{-10}$\\
Initial learning rate $L_{r}$ & $0.1$ \\
Decaying rate $d_{r}$ & $0.95$ \\
Decaying epoch $d_{e}$ & $100$ \\
$L_{0.5}$ trasition point $a$ & $0.01$ \\
Threshold scale $b$ & $0.05$ \\
\specialrule{0.05em}{2pt}{0pt}
\end{tabular}
\end{table}

\section{\label{sec:results}Experiments and Results}
In this section, we test the performance of the GEPNN using a series of cases. First, the GEPNN is applied to identify the algebraic equations of three different physical laws, including the law of ideal gas, the law of universal gravitation, and the law of the wall in a turbulent channel flow. These cases test the GEPNN’s capabilities, including the convergence performance, accuracy of constant numbers in complex situations, and robustness under perturbation of the training data. Next, GEPNN is applied to modeling the subgrid scale (SGS) stress for large-eddy simulation (LES) of homogeneous isotropic turbulence. The GEPNN model is then evaluated in \emph{a priori} and \emph{a posteriori} tests. 

\subsection{\label{subsec:laws}Training cases of physics laws}
In this subsection, we test the GEPNN method by training three physics laws based on synthetic data. 
The first case, the ideal gas law, is basic and consists of two input variables. The second case (\emph{i.e.}, the gravitation law) is more complex, involving three input variables and requiring an additional data-normalization operation. The third case, the logarithmic law in a turbulent channel flow includes various degrees of noise, testing the robustness of the GEPNN. Table~\ref{tab:caseset} summarizes the setup and parameters of these three cases. 

\begin{table}[]
\centering\small
\caption{\label{tab:caseset}Setup and parameters of physics laws}
\begin{tabular}{cccc}
\specialrule{0.03em}{1pt}{0pt}
Case No. & Laws & Input variables & Noise \\\specialrule{0.03em}{1pt}{1pt}
1 & Ideal gas law & $R, T$ & $0\%$ \\
2 & Gravitation law & $m_{1}, m_{2}, 1/r$ & $0\%$ \\
3 & Logarithmic law & $\ln y^{+}$ & $0\%, \pm20\%, \pm80\%$ \\\specialrule{0.03em}{1pt}{0pt}
\end{tabular}
\end{table}


\subsubsection{\label{subsubsec:idealgas}The law of ideal gas}
In the first case, the GEPNN is used to identify the equation for the ideal gas law and predict the universal gas constant~\cite{2005Student}. This law describes the relationship between pressure $p$, volume $V$, amount of material $n$, and temperature $T$ when the ideal gas is in equilibrium. The equation of state of the ideal gas can be expressed as follows:
\begin{equation}
\label{eq:idealgas}
pV=nRT,
\end{equation}
where $R=8.314J\cdot mol^{-1}K^{-1}$ is the universal gas constant obtained using the international system of units. Assuming that $n$ and $T$ are two input variables and $pV$ is the output, we can use the GEP and GEPNN to identify this expression and obtain the universal gas constant $R$. The training dataset is generated using $0\leqslant n \leqslant 1$ and $0\leqslant T\leqslant 1$, and $pV$ is calculated using Eqn.~(\ref{eq:idealgas}). 

\begin{figure}
\centering\footnotesize
\begin{overpic}[scale=0.35]{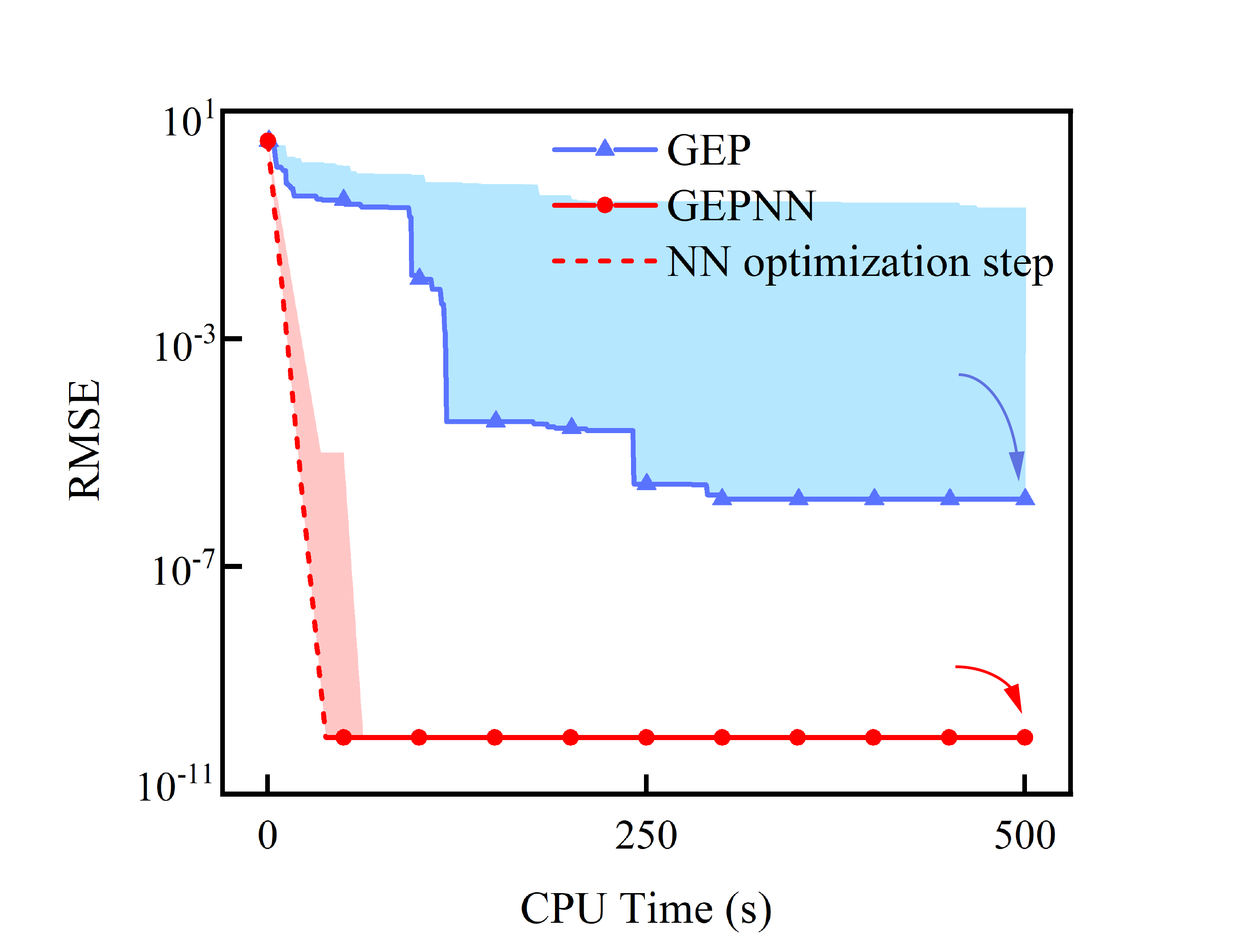}
\put(40,45){\textcolor[RGB]{94,114,255}{$8.312nT+0.0007n+0.0007T$}}
\put(66,22){\textcolor[RGB]{255,0,0}{$8.314nT$}}
\end{overpic}
\caption{Convergence comparison of the ideal gas law. The randomness was subtracted from the RMSE for normalization. 
The red dashed line refers to the neural network optimization step in the GEPNN. The blue and red areas correspond to ten training sessions, and the solid lines denote the best result for each method.}
\label{fig:gepnn_states}
\end{figure}

The training progress of GEP and the GEPNN are given in Fig.~\ref{fig:gepnn_states}, illustrating that the GEPNN converges directly to the optimum, while GEP converges to a higher RMSE than the GEPNN.
The resulting expressions of the GEP and GEPNN are as follows:
\begin{equation}
\begin{aligned}
\label{eq:ideal_gas_result}
\text{GEP}:&pV=8.312nT+0.0007n + 0.0007T,\\
\text{GEPNN}:&pV=8.314nT.
\end{aligned}
\end{equation}
After the generation which applies the SRNN optimization, the RMSE of the GEPNN noticeably descends to the convergence threshold, as shown by the red dashed line in Fig.~\ref{fig:gepnn_states}. With the SRNN optimization step, the constant number before $nT$ approaches the target value, and the constants before the other ineffective terms are reduced to $0$. Thus, the GEPNN can identify the correct format of the state equation of ideal gas with the exact universal gas constant $R^{GEPNN}=8.314$, whereas GEP produces additional terms with incorrect coefficients, which leads to a larger RMSE than GEPNN. 

These results reflect the properties of symbolic regression tasks discussed in Sec.~\ref{subsec:GEPvsSRNN}. Traditional GEP has the advantage of producing various expression structures. However, without suitable coefficients, the correct expression structure cannot provide a low RMSE. Therefore, in this case, GEP provides the results with additional terms to fit the data. In comparison, GEPNN performs efficiently in both parts, providing a correct expression form with an appropriate coefficient. 

\subsubsection{\label{subsubsec:gravity}The law of universal gravitation}
In this case, the GEPNN is used to discover the universal gravitation law with the gravitational constant to verify the convergence ability of the GEPNN in a more complex situation involving three variables. 

The law of universal gravitation~\cite{Newton2000The}, discovered by Isaac Newton, describes gravity as the attractive force between two objects. Gravity obeys the following equation:
\begin{equation}
\label{eq:gravity}
F=G\frac{m_{1}m_{2}}{r^{2}},
\end{equation}
where $m_{1}$ and $m_{2}$ are the masses of two objects, $r$ is the distance between two objects, and $G=6.67\times 10^{-11}N\cdot m^{2}kg^{-2}$ is the gravitational constant. Observing the inverse relationship between gravity and distance, the input variables are selected as $m_{1}$, $m_{2}$, and $1/r$.

In this case, training the gravitational constant is not straightforward, as it is on the order of $10^{-11}$, which needs a higher precision data type and is not conducive to data processing and display. Therefore, we must generate a training dataset with normalization to obtain a more effective target by manipulating the data as
\begin{equation}
\label{eq:norm_gravity}
F=(G \times 10^{10}) \frac{(m_{1} / 10^{5})(m_{2} / 10^{5})}{r^{2}} =G^{*}\frac{m_{1}^{*}m_{2}^{*}}{r^{2}},
\end{equation}
where $G^{*}=0.667$ is the normalized gravitational constant and $m_{1}^{*}$ and $m_{2}^{*}$ are normalized masses. 
The training dataset is generated using $0 \leqslant m_{1}^{*},m_{2}^{*} \leqslant 1$ and $0 < r < 1$, and gravity is calculated with Eqn.~(\ref{eq:norm_gravity}).

\begin{figure}
\centering\footnotesize
\begin{overpic}[scale=0.35]{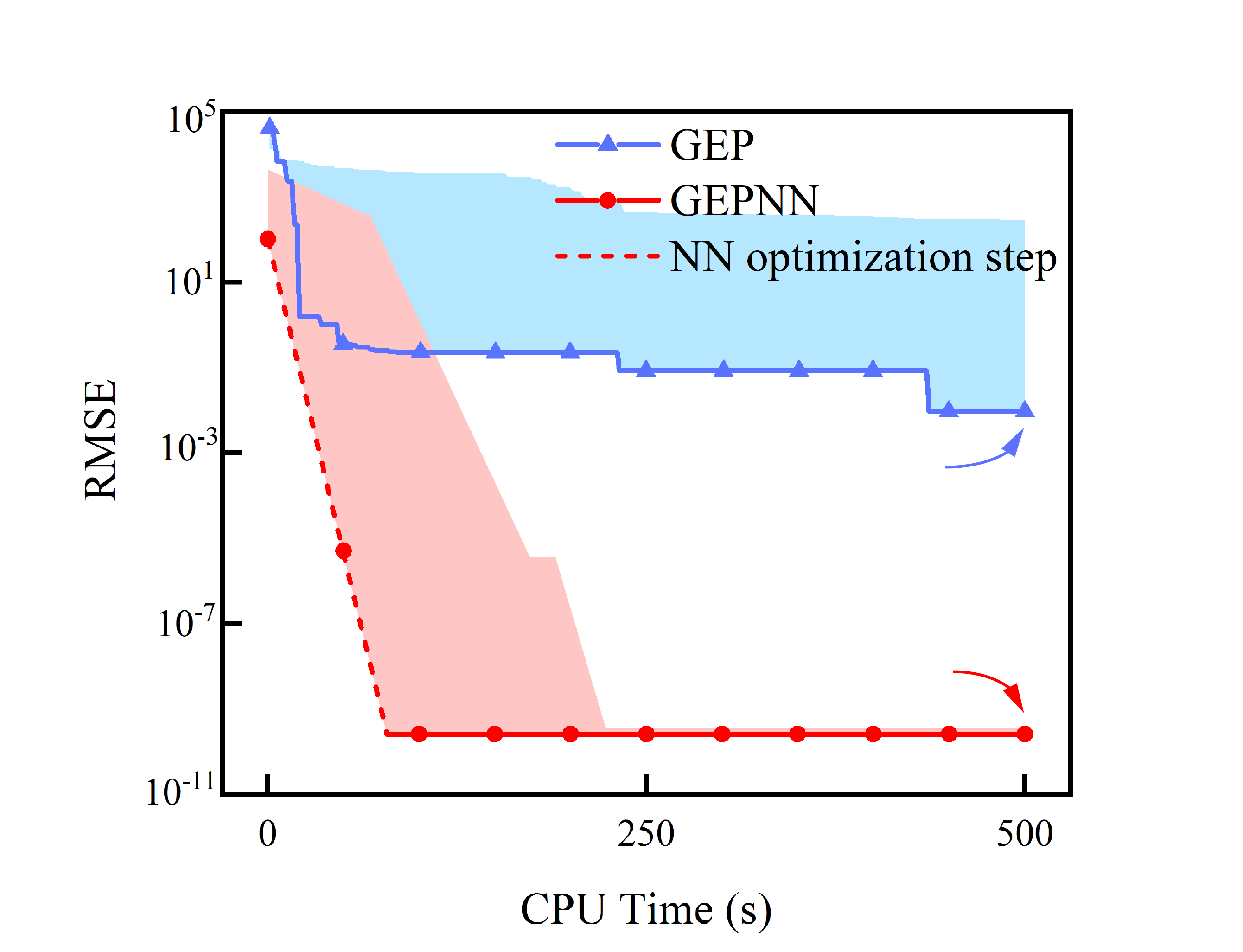}
\put(36,38){\textcolor[RGB]{94,114,255}{$0.672\frac{m_{1}^{*} m_{2}^{*}}{r^{2}}+0.024\frac{m_{1}^{*} m_{2}^{*}}{r}+0.022$}}
\put(63,22){\textcolor[RGB]{255,0,0}{$0.667\frac{m_{1}^{*} m_{2}^{*}}{r^{2}}$}}
\end{overpic}
\caption{Convergence comparison for the law of universal gravitation. 
The blue and red areas show the results from ten training sessions, and the solid lines indicate the best results from each method. The red dashed line denotes the ANN optimization step in the GEPNN.}
\label{fig:gepnn_gravity}
\end{figure}

The training progress on the gravitational law data with a normalized gravitational constant are presented in Fig.~\ref{fig:gepnn_gravity}. The final expressions from the GEP and GEPNN are as follows:
\begin{equation}
\begin{aligned}
\text{GEP}:&F=0.672\frac{m_{1}^{*}m_{2}^{*}}{r^{2}}+0.024\frac{m_{1}^{*}m_{2}^{*}}{r}+0.022,\\
\text{GEPNN}:&F=0.667\frac{m_{1}^{*}m_{2}^{*}}{r^{2}}
\end{aligned}
\end{equation}
Both the GEP and GEPNN provide the primary format $m_{1}^{*}m_{2}^{*}/r^{2}$ of the universal gravitation law. The GEPNN accurately identifies the constant with the optimization of the SRNN, whereas GEP cannot precisely determine the constant which results in unnecessary terms. 

\subsubsection{\label{subsubsec:loglaw}The law of  wall in turbulent channel flow}
In this subsection, artificially generated turbulent channel flow velocity data with certain fluctuations are used to test the robustness of the GEPNN. 

In high Reynolds number turbulent channel flows, there is an inner layer close to the wall in which the mean velocity profile is determined by viscous scales~\cite{pope2000turbulent}. Using the dimensionless variables $y^{+}$ and $u^{+}$, which are defined by
\begin{equation}
y^{+}\equiv y/\delta_{\nu},~ u^{+}\equiv \langle U\rangle /u_{\tau},
\end{equation}
the law of the wall can be expressed as
\begin{equation}
u^{+}=f_{w}(y^{+}),
\end{equation}
where $y$ denotes the wall-normal distance, $\delta_{\nu}$ denotes the viscous length scale, $\langle U\rangle$ represents the mean velocity, $u_{\tau}$ denotes the friction velocity, and $f_{w}(*)$ denotes the wall function. Additionally, $u^{+}$ and $y^{+}$ represent the velocity and wall-normal distance in the wall units, respectively.

In general, in the region where $y^{+}>30$, the following law exists for $y^{+}$ and $u^{+}$:
\begin{equation}
\label{eq:loglaw}
u^{+}=\frac{1}{\kappa} \ln y^{+}+B.
\end{equation}
This equation is the logarithmic law of the wall (log-law), and the region where $u^{+}$ and $y^{+}$ obey this equation is called the log-law region. The constant numbers in the log-law vary to some extent, but they are usually within $5\%$ of 
\begin{equation}
\label{eq:loglaw_const}
\kappa = 0.41, B=5.2.
\end{equation}

By analyzing the logarithmic relationship between $y^{+}$ and $u^{+}$, we use $\ln y^{+}$ and $u^{+}$ as input and output variables, respectively, and generate training data with Eqn.~(\ref{eq:loglaw}) and Eqn.~(\ref{eq:loglaw_const}), where $30\leqslant y^{+} \leqslant 100$. To evaluate the tolerance of data randomness for the GEPNN, three sets of data are created with no randomness, $\pm20\%$ and $\pm80\%$ randomness of $u^{+}$, denoted as cases A, B, and C. 
The randomness is assigned as $u^{+}=[\frac{1}{\kappa} \ln y^{+}+B](1\pm r_{i})$, where $r_{i}$ is a random number sampled from a uniform distribution between $-0.2$ to $0.2$ and $-0.8$ to $0.8$, respectively, and the average value of the data is consistent with that generated without randomness. 

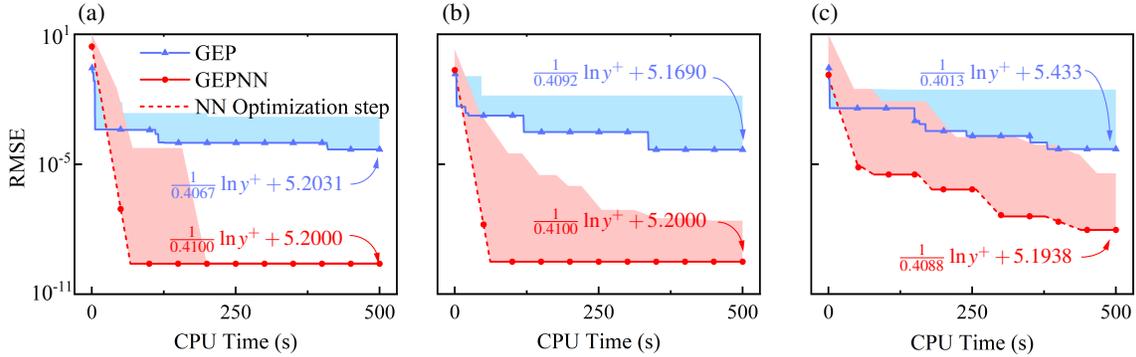
\begin{figure}[!ht]
\centering\footnotesize
\hspace*{-1.8cm}
\begin{overpic}[scale=0.23]{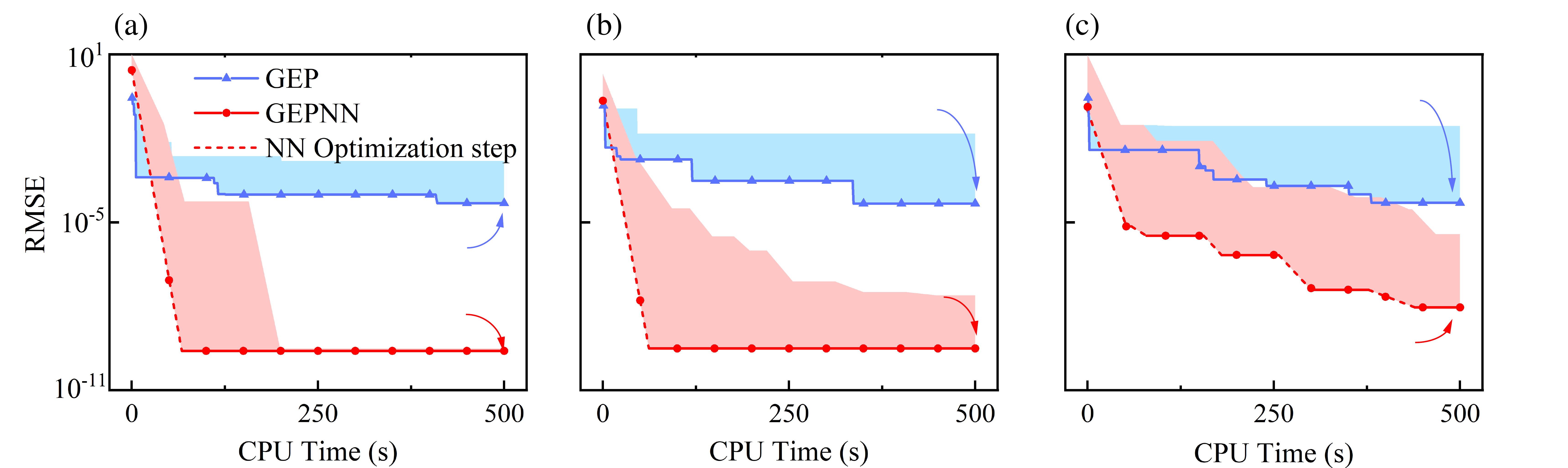}
\put(14.7,13.5){\textcolor[RGB]{94,114,255}{$\frac{1}{0.4067}\ln{y^{+}}+5.2031$}}
\put(14.6,9){\textcolor[RGB]{255,0,0}{$\frac{1}{0.4100}\ln{y^{+}}+5.2000$}}
\put(44.7,22.8){\textcolor[RGB]{94,114,255}{$\frac{1}{0.4092}\ln{y^{+}}+5.1690$}}
\put(44.7,10.5){\textcolor[RGB]{255,0,0}{$\frac{1}{0.4100}\ln{y^{+}}+5.2000$}}
\put(76.7,22.8){\textcolor[RGB]{94,114,255}{$\frac{1}{0.4013}\ln{y^{+}}+5.433$}}
\put(75,7.5){\textcolor[RGB]{255,0,0}{$\frac{1}{0.4088}\ln{y^{+}}+5.1938$}}
\end{overpic}
\caption{Convergence comparison for (a) case A, (b) case B, and (c) case C. The randomness has been subtracted from the RMSE of cases B and C for normalization. The blue and red areas show the results from ten training sessions, and the solid lines refer to the best result for each method. The red dashed line refers to the neural network optimization step in the GEPNN.}
\label{fig:gepnn_loglaw}
\end{figure}

Fig.~\ref{fig:gepnn_loglaw} shows the training progress for these three cases. 
The GEP and GEPNN obtain a mathematical expression with a linear relationship between $\ln y^{+}$ and $u^{+}$, but the slopes $\kappa$ and intercept $B$ are different. Table~\ref{tab:loglaw_gepnn} lists the constant numbers for each case. In all cases, the GEPNN converges faster than the GEP, and the former achieves a smaller RMSE than the latter. With increasing randomness of the data, the convergence RMSEs of the GEP and GEPNN worsen, especially for GEPNN. Also, the convergence time consumed by GEPNN becomes larger with the increase of randomness. However, the GEPNN provides higher precision for $\kappa$ (two digits after the decimal point) and $B$ (one digit after the decimal point), proving that GEPNN is robust and can tolerate a certain amount of randomness and still generate the optimum result.
\begin{table}[]
\centering\small
\caption{\label{tab:loglaw_gepnn}Constant numbers $\kappa$ and $B$ for cases A, B, and C}
\begin{tabular}{lcccccc}
\specialrule{0.03em}{1pt}{0pt}
& \multicolumn{2}{c}{case A} & \multicolumn{2}{c}{case B} & \multicolumn{2}{c}{case C} \\\specialrule{0.03em}{2pt}{2pt}
Constant & $\kappa$ & b & $\kappa$ & b & $\kappa$ & b \\
\multicolumn{1}{c}{GEP} & 0.4067 & 5.2031 & 0.4092 & 5.169 & 0.4013 & 5.433 \\
\multicolumn{1}{c}{GEPNN} & 0.4100 & 5.2000 & 0.4100 & 5.2000 & 0.4088 & 5.1938 \\\specialrule{0.03em}{2pt}{0pt}
\end{tabular}
\end{table}

\subsection{\label{subsec:SGS}Applying the GEPNN to modelling subgrid-scale stress in large-eddy simulation}
In this section, we apply the GEPNN to subgrid scale (SGS) stress modelling in large-eddy simulation (LES). 
LES is an important method for the numerical simulation of turbulence and has been applied widely in various applications in, for example, automobile design and aerospace engineering. By filtering the governing equations, the LES solves the large-scale flow field and models the effects of SGS structures on the large-scale flow field. These effects are considered by modelling the stress term $\tau_{ij}$ in the filtered governing equations, and this type of model is called SGS closure. Accurate modelling of the SGS stress is essential for turbulence simulation to predict the precise effects of subgrid-scale structures. With the development of LES, different models have been developed, including the Smagorinsky model~\cite{smagorinsky1963general}, the dynamic Smagorinsky model (DSM)~\cite{moin1991dynamic}, the dynamic mixed model (DMM)~\cite{liu1994properties}, and the gradient model~\cite{clark1979evaluation}, \emph{etc}. 
Although these models have been tested under different flow conditions and shown some success in predicting mean flow behavior, the physical quantities predicted by these models sometimes have low correlations with those of the DNS results. Using data-driven approaches to model the SGS stress aims to predict physical quantities more consistent with the DNS data. 

The governing equations for 3-D incompressible turbulence~\cite{pope2000turbulent, buzzicotti2018effect} are as follows:
\begin{equation}
\begin{aligned}
\label{eq:ns}
\frac{\partial u_{i}}{\partial x_{i}} &= 0,\\
\frac{\partial u_{i}}{\partial t} + \frac{\partial u_{i}u_{j}}{\partial x_{j}}
= -& \frac{\partial p}{\partial x_{i}} + 
\nu\frac{\partial^{2}u_{i}}{\partial x_{j}x_{j}} + \mathcal{F}_{i},
\end{aligned}
\end{equation}
where $u_{i}$ is the velocity component in the $i^{th}$ direction, $p$ is the pressure, $\nu$ is the kinematic viscosity, and $\mathcal{F}_{i}$ is a large-scale forcing.

In LES, the governing equations are filtered to separate large-scale motion and SGS stress. By applying the filter operation to Eqn.~(\ref{eq:ns}), the filtered governing equations~\cite{meneveau2000scale} become
\begin{equation}
\begin{aligned}
\label{eq:f_continous}
\frac{\partial{\widetilde{u}}_{i}}{\partial x_{i}} &= 0,\\
\frac{\partial{\widetilde{u}}_{i}}{\partial t} + 
\frac{\partial{\widetilde{u}}_{i}{\widetilde{u}}_{j}}{\partial x_{j}} = 
- \frac{\partial\widetilde{p}}{\partial x_{i}} &-
\frac{\partial\tau_{ij}}{\partial x_{j}} + 
\nu\frac{\partial^{2}{\widetilde{u}}_{i}}{\partial x_{j}x_{j}} + 
{\widetilde{\mathcal{F}}}_{i},
\end{aligned}
\end{equation}
where the overbar variable $\widetilde{*}$ is referring to filtered quantities, and the SGS stress tensor $\tau_{ij}$ can be expressed as
\begin{equation}
\tau_{ij} = \widetilde{u_{i}u_{j}} - {\widetilde{u}}_{i}{\widetilde{u}}_{j}.
\end{equation}
This term is not closed because $\widetilde{u_{i}u_{j}}$ cannot be expressed with quantities from the filtered equations. Therefore, the SGS stress tensor $\tau_{ij}$ has to be modelled using known quantities from the flow field. 
According to the Cayley–Hamilton theory following Pope~\cite{pope1975more}, the anisotropic part of the SGS stress tensor $\tau_{ij}^{A}$ can be expressed as
\begin{equation}
\label{eq:modelexpression}
\tau_{ij}^{A} = {\sum\limits_{n = 1}^{10}{g^{(n)}\left( {I_{1},\ldots,I_{4}} \right)T^{(n)}}},
\end{equation}
where $T^{(n)}$ is a tensor basis function, ${\widetilde{S}}_{ij}$ is the filtered strain-rate tensor, ${\widetilde{\Omega}}_{ij}$ is the rotation-rate tensor, and $I_{i}$ are independent invariants. The coefficients $g^{(n)}$ are functions of the invariants, which must be predicted using data-driven approaches. 

The filtered strain rate tensor ${\widetilde{S}}_{ij}$ and the rotation rate tensor ${\widetilde{\Omega}}_{ij}$ are defined as follows:
\begin{equation}
{\widetilde{S}}_{ij} = 
\frac{1}{2}\left( {\frac{\partial{\widetilde{u}}_{i}}{\partial x_{j}} + 
\frac{\partial{\widetilde{u}}_{j}}{\partial x_{i}}} \right),
~{\widetilde{\Omega}}_{ij} = 
\frac{1}{2}\left( {\frac{\partial{\widetilde{u}}_{i}}{\partial x_{j}} - 
\frac{\partial{\widetilde{u}}_{j}}{\partial x_{i}}} \right).
\end{equation}
Using the inverse timescale $|\widetilde{S}|$ to nondimensionalise ${\widetilde{S}}_{ij}$ and ${\widetilde{\Omega}}_{ij}$, the dimensionless quantities $s_{ij}$ and $\omega_{ij}$ are used to build the SGS model
\begin{equation}
|\widetilde{S}| = \sqrt{{\widetilde{S}}_{mn}{\widetilde{S}}_{mn}},
~s_{ij} = \frac{{\widetilde{S}}_{ij}}{|\widetilde{S}|},
~\omega_{ij} = \frac{{\widetilde{\Omega}}_{ij}}{|\widetilde{S}|}.
\end{equation}
The first four normalized tensor basis functions $T^{(1\dots 4)}$ and the first four normalized invariants $I_{1\dots 4}$ are selected: 
\begin{alignat}{2}
\label{eq:tensor_basis}
&T^{(1)}=s,&&T^{(2)}=s\omega-\omega s,\nonumber\\
&T^{(3)}=s^{2}-\frac{1}{3}I\cdot Tr(s^{2}),&&T^{(4)}= \omega^{2}-\frac{1}{3}I\cdot Tr(\omega^{2}),\\
&I_{1}=Tr(s^{2}), I_{2}=Tr(\omega^{2}),&&I_{3}=Tr(s^{3}),I_{4}=Tr(\omega^{2}s),\nonumber
\end{alignat}
where $I$ denotes an identity matrix. For further details on the training setup, refer to ~\cite{li2021data}. 

\begin{figure}[!ht]
\centering
\hspace*{-0.9cm}
\includegraphics[width=15cm]{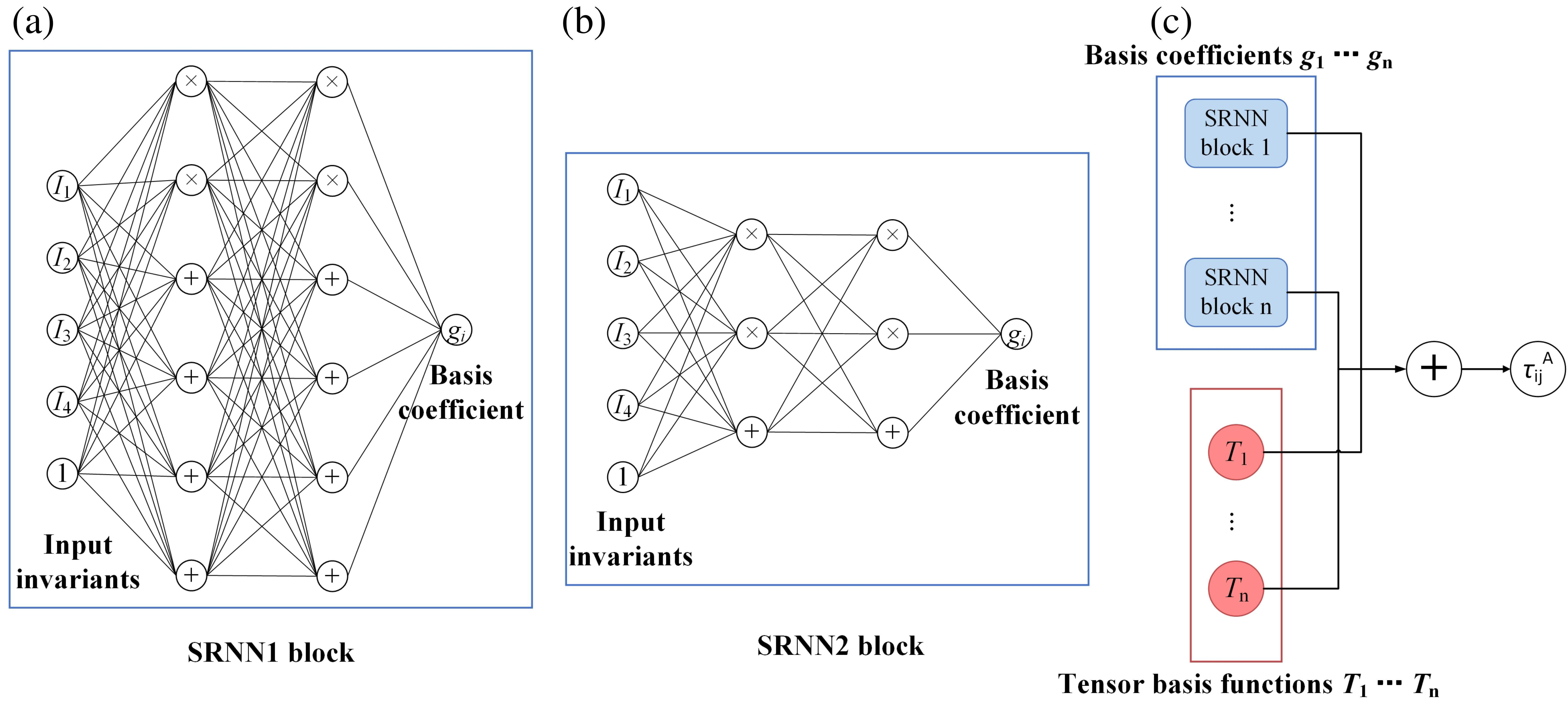}
\caption{Architectures of SRNN and modelling strategies. (a) SRNN1 with a complex structure. (b) SRNN2 with a simpler structure. (c) SGS stress modelling with SRNN.}
\label{fig:sgs_srnn}
\end{figure}

To compare the differences between GEP, SRNN, and GEPNN, two SRNNs with different architectures, shown in Fig.~\ref{fig:sgs_srnn}(a) and Fig.~\ref{fig:sgs_srnn}(b), are designed and used for SGS stress modelling. Different from the SRNNs in Sec.~\ref{subsec:GEPvsSRNN}, four separated SRNN blocks are used to predict the four basic coefficients $g^{(1\dots 4)}$. The SRNN blocks are connected with their tensor basis functions to generate the SGS stress and the gradient can be backpropagated to each block to ensure that all the blocks can be trained simultaneously.

The training dataset is extracted from the homogeneous isotropic turbulence~\cite{xie2020artificial, li2021data}. The grid size of the DNS is $1,024^{3}$, and the Taylor Reynolds number is $Re_{\lambda}\approx 260$~\cite{Wang2018KineticET}. The filtered velocity, corresponding SGS stress $\tau_{ij}$, and other filtered variables are obtained using a top-hat filter with a width of $\mathrm{\Delta} = 16 \mathrm{\Delta} x$. Using all the data points in the filtered DNS (fDNS) data for training is inefficient and redundant, thus we downsampled the fDNS data and selected about $2\times 10^{6}$ pairs of data points for the training process.

During the training stage, the correlation coefficient $C(\tau)$ of the DNS SGS stress $\tau$ and the modelled SGS stress $\tau^{model}$ are used as \emph{a priori} tests to evaluate the performance of the SGS models. $C(\tau)$ is defined as
\begin{equation}
C\left( \tau \right) = \frac{\left\langle {\left\langle {\tau - \left\langle \tau \right\rangle} \right\rangle\left\langle {\tau^{model} - \left\langle \tau^{model} \right\rangle} \right\rangle} \right\rangle}{\left( {\left\langle \left\langle {\tau - \left\langle \tau \right\rangle} \right\rangle^{2} \right\rangle\left\langle \left\langle {\tau^{model} - \left\langle \tau^{model} \right\rangle} \right\rangle^{2} \right\rangle} \right)^{1/2}},
\end{equation}
where $\left\langle {~ * ~} \right\rangle$ denotes the averaging over the spatial volume. 

\begin{figure}[!ht]
\centering
\hspace*{-1.4cm}
\includegraphics[width=15cm]{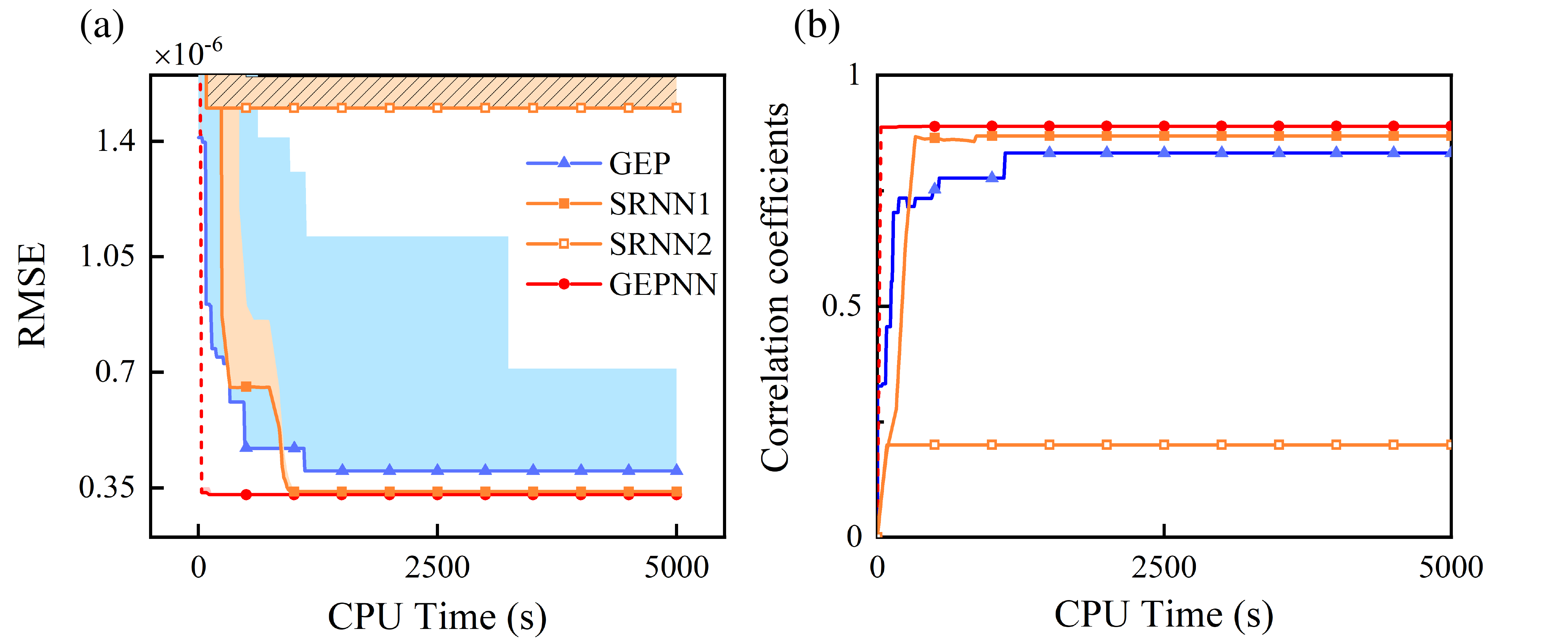}
\caption{Modeling SGS stress tensor $\tau_{ij}$ using the GEP, SRNN, and GEPNN methods. For the GEPNN plots, the red dashed parts represent the SRNN optimization step in the training iterations. The lines denote the best performance among $30$ training sessions for each method, while the shaded areas indicate variations between different sessions. (a) Convergence comparison for training SGS models. (b) Comparison of correlation coefficients.}
\label{fig:gepnn_sgs_fitness}
\end{figure}

Fig.~\ref{fig:gepnn_sgs_fitness} presents the RMSE and correlation coefficients of the GEP, SRNN, and GEPNN SGS models during training progress. 
For each method, we run $30$ training sessions, and the lines in Fig.~\ref{fig:gepnn_sgs_fitness}(a) represents the best results for each method, while the shaded areas indicate variations among different training sessions.
It is firstly noted that it is quite difficult for the SRNN2 with its simple structure to converge to a reasonable result, even using the same hyperparameters as for the SRNN1. 
This illustrates the limitation of the inflexible preset structure of SRNN. 
Furthermore, the RMSE and correlation coefficients obtained by the SRNN1 and GEPNN models are both very promising, better than the traditional GEP model, indicating that SRNN and GEPNN have better convergence performance given the same training time.
Moreover, the GEPNN performs slightly better compared to the SRNN1, especially for the correlation coefficients shown in Fig.~\ref{fig:gepnn_sgs_fitness}(b).


The final expressions of the GEP, SRNN1 and GEPNN models, which are selected as the best one from the $30$ training sessions for each method, are as follows:
\begin{equation}
\begin{aligned}
\label{eq:sgs_result}
\text{GEP}&:\tau=( {\mathrm{\Delta}| \widetilde{S} |} )^{2}\left( {- 0.1000T^{(2)} + 0.1130T^{(3)} - 0.1015T^{(4)}} \right),\\
\text{SRNN1}&:\tau=( {\mathrm{\Delta}| \widetilde{S} |} )^{2}\left( {- 0.1055T^{(2)} - 0.1042T^{(4)}} \right),\\
\text{GEPNN}&:\tau=( {\mathrm{\Delta}| \widetilde{S} |} )^{2}\left( {- 0.0123T^{(1)} - 0.1043T^{(2)} + 0.0701T^{(3)} - 0.1039T^{(4)}} \right).
\end{aligned}
\end{equation}
It is noted that $g^{(1\dots 4)}$ contain no invariants $I_{1\dots 4}$, and all terms of $g^{(1\dots 4)}$ are constants. 
Table~\ref{tab:gepnn_sgs_statistic} summarises the constant coefficients $g^{(1\dots 4)}$ for the GEP, SRNN1 and GEPNN models obtained from $30$ training sessions. The optimization process of GEP includes a series of random processes, for example, the probability of mutation and crossover. Thus, the convergence results of the GEP method may differ for each training run within a limited time. 
As a comparison, the GEPNN shows good convergence within a limited training time because the neural network optimization step is very efficient with gradient information to obtain the correct optimization direction.
Among the $30$ training sessions of SRNN1, however, the trained coefficients fucntions $g^{(1)}$ and $g^{(3)}$ are either $0$, or complex combinations of invariants $I_{1\dots 4}$ with coefficients of small magnitudes (not shown in Table~\ref{tab:gepnn_sgs_statistic}).
This suggests that the strategies we introduced to enhance the model sparsity in GEPNN do result in better convergence to simplified models compared to the original SRNN. 
 
\begin{table}[]
\centering\small
\caption{\label{tab:gepnn_sgs_statistic}Summary of constant coefficients of the GEP and GEPNN models obtained from $30$ training processes.}
\begin{tabular}{lccc}
\specialrule{0.03em}{3pt}{3pt}
& \multicolumn{1}{c}{GEP} & \multicolumn{1}{c}{SRNN1} & \multicolumn{1}{c}{GEPNN}\\\specialrule{0.03em}{2pt}{2pt}
$g^{(1)}$ & $-0.0100 \pm 0.0100$ &  $0 \pm 0$  & $-0.0119 \pm 0.0014$ \\
\multicolumn{1}{c}{$g^{(2)}$} &$-0.1025 \pm 0.0075$ &  $-0.1055 \pm 0.0001$  &  $-0.1047 \pm 0.0040$ \\
\multicolumn{1}{c}{$g^{(3)}$} &$0.0800 \pm 0.0200$ &  $0 \pm 0$  &  $0.0699 \pm 0.0011$ \\
\multicolumn{1}{c}{$g^{(4)}$} &$-0.1080 \pm 0.0070$ &  $-0.1042 \pm 0.0003$  &  $-0.1031 \pm 0.0014$ \\\specialrule{0.03em}{2pt}{0pt}
\end{tabular}
\end{table}

\begin{figure}[!ht]
\centering
\includegraphics[width=10cm]{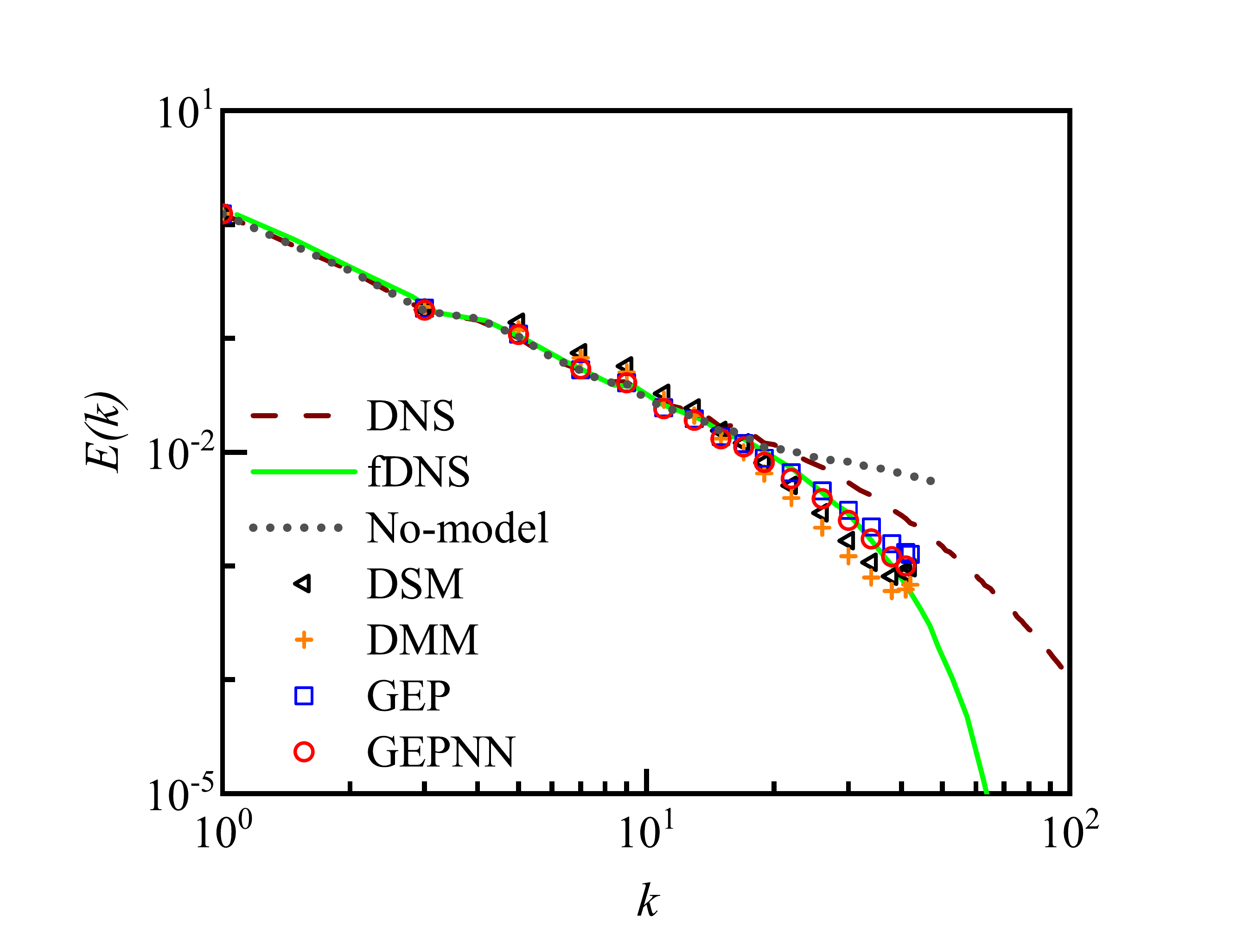}
\caption{Velocity spectrum for LES using different models at a grid resolution of $128^{3}$. DNS refers to the DNS data of $1,024^{3}$, while fDNS refers to the DNS data filtered using $\mathrm{\Delta} = 16 \mathrm{\Delta} x$. No-model is the LES without SGS models.}
\label{fig:sgs_ek}
\end{figure}

Furthermore, we apply the trained GEP and GEPNN models to LES of homogeneous isotropic turbulence at a grid resolution of $128^{3}$, and the \textit{a posteriori} results from these ML models are then investigated. 
Moreover, the traditional SGS models, \emph{i. e.} DSM and DMM, are also applied and shown for comparison. 
Fig.~\ref{fig:sgs_ek} shows the spectrum of the velocity field. The velocity spectrum of the DNS has a long inertial region. Owing to the filtering operation, the fDNS case deviates in the high-wavenumber region. For comparison, we conducted an LES without SGS models at a grid resolution of $128^{3}$, \emph{i.e.} an under-resolved DNS. A significant deviation occurred near the cutoff wavenumber because the simulation failed to account for the dissipation of the unresolved scales. The spectra of the DSM and DMM models, however, are damped near the cutoff wavenumber but are energy-rich at lower wavenumbers. The GEP and GEPNN models predict velocity spectra that nearly overlap with the fDNS data, with the GEPNN model being more accurate near the cutoff wavenumber. 

\begin{figure}[!ht]
\centering
\includegraphics[width=10cm]{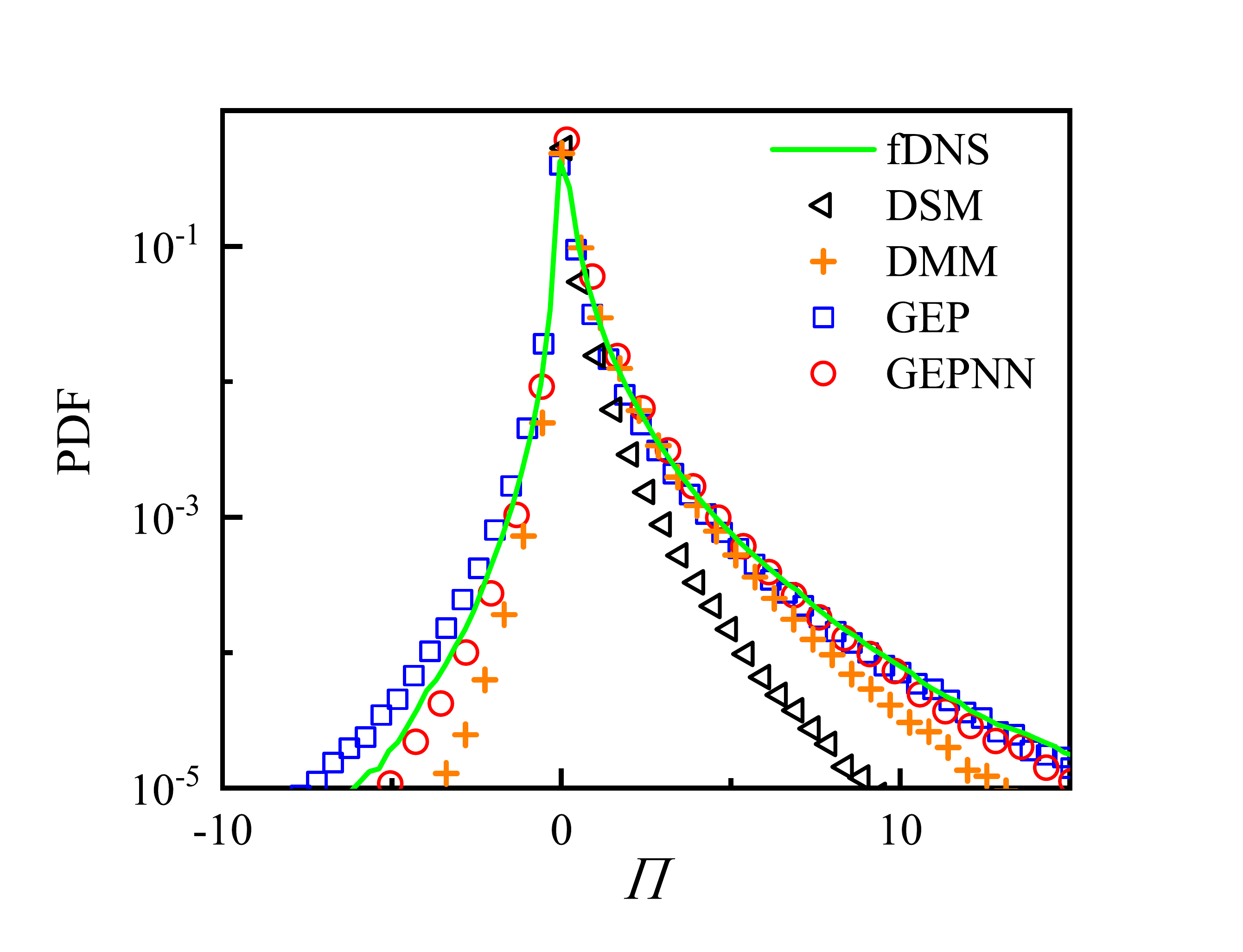}
\caption{PDFs of SGS energy flux for fDNS and LES using different SGS models at a grid resolution of $128^{3}$.}
\label{fig:sgs_flux}
\end{figure}

The SGS energy flux is also evaluated to investigate the energy transfer between the resolved and subgrid scales, which is defined as the product of the SGS stress tensor and the strain rate tensor
\begin{equation}
\Pi=-\tau_{ij} \widetilde{S}_{ij}.
\end{equation}
Fig.~\ref{fig:sgs_flux} shows the PDFs of the SGS energy flux $\Pi$ for the LES using different models, where the positive SGS flux refers to the energy transfer from the resolved field to the subgrid field, and the negative flux is the energy transfer in the opposite direction. Compared to the DSM and DMM models, the GEP model predicts a positive SGS flux similar to the fDNS data, but overpredicts the negative flux. However, the GEPNN model makes a precise prediction, not only for the forward energy transfer but also for the backscatter of kinetic energy from the subgrid field to the resolved scale. 

\begin{figure}[!ht]
\centering
\hspace*{-1.2cm}
\includegraphics[width=14cm]{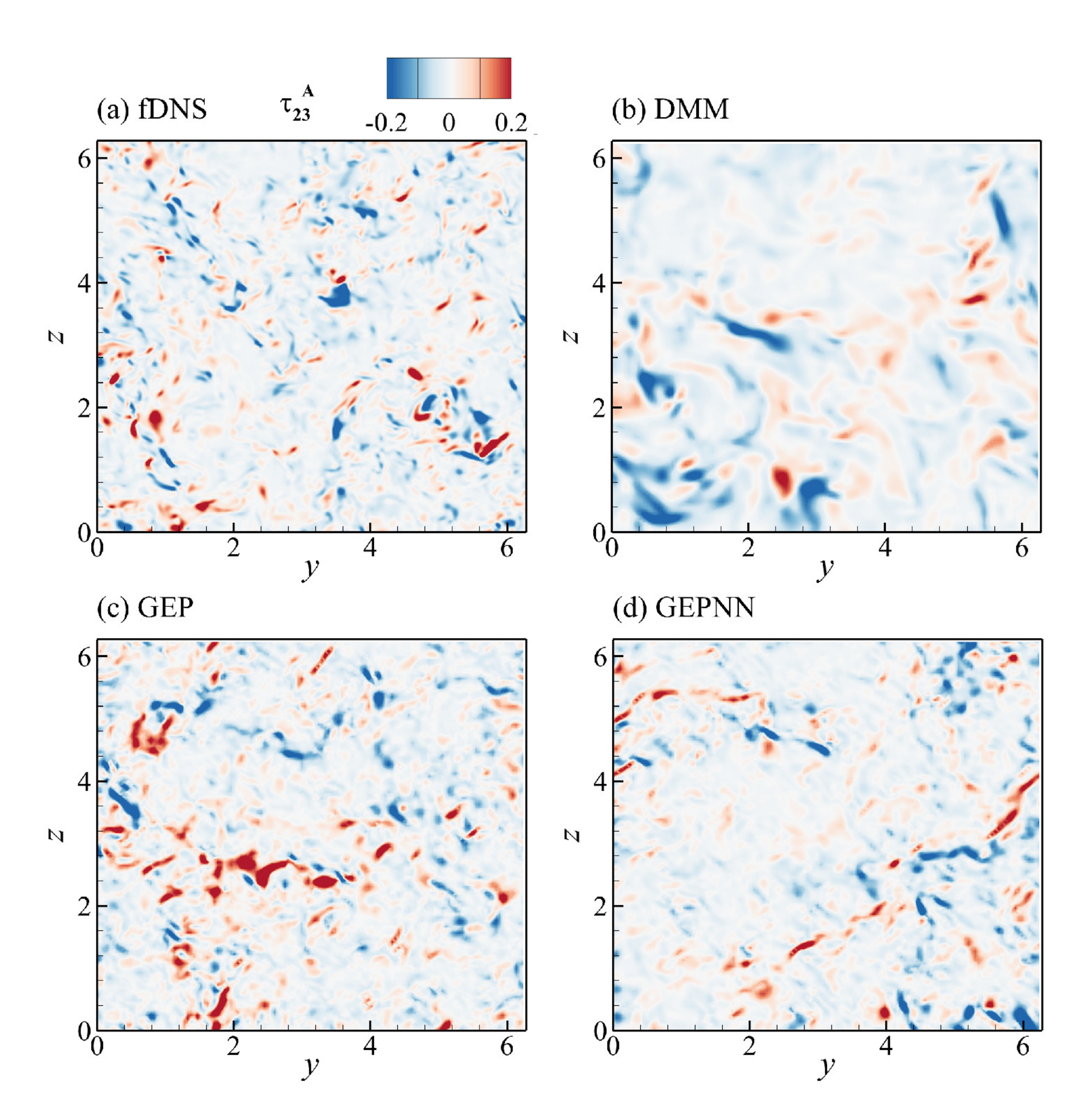}
\caption{Contours of the SGS stress $\tau_{23}^{A}$ for the fDNS data and LES at an arbitrary slice at grid resolution of $128^{3}$ using different SGS stress models: (a) fDNS, (b) DMM, (c) GEP, (d) GEPNN.}
\label{fig:sgs_tau23}
\end{figure}

The contours of one component of the SGS stress tensor $\tau_{ij}^{A}$ at an arbitrary slice of the fDNS and LES using different SGS models are shown in Fig.~\ref{fig:sgs_tau23}, where regions with high stress intensities represent localized subgrid stress structures. The structures generated by the DMM model are larger and more diffusive than the fDNS data and those generated by data-driven models, because the DMM introduces extra dissipation that dissipates small-scale structures.
However, the small-scale structures can be effectively captured using the GEP and GEPNN models, similar to the fDNS data, illustrating that data-driven closures of SGS stress can model small-scale flow behaviour better than the DMM model. This observation is consistent with the previous findings and demonstrates the superior performance of the data-driven SGS models, as illustrated in Fig.~\ref{fig:sgs_ek} and Fig.~\ref{fig:sgs_flux}.

\begin{figure}[!ht]
\centering
\includegraphics[width=10cm]{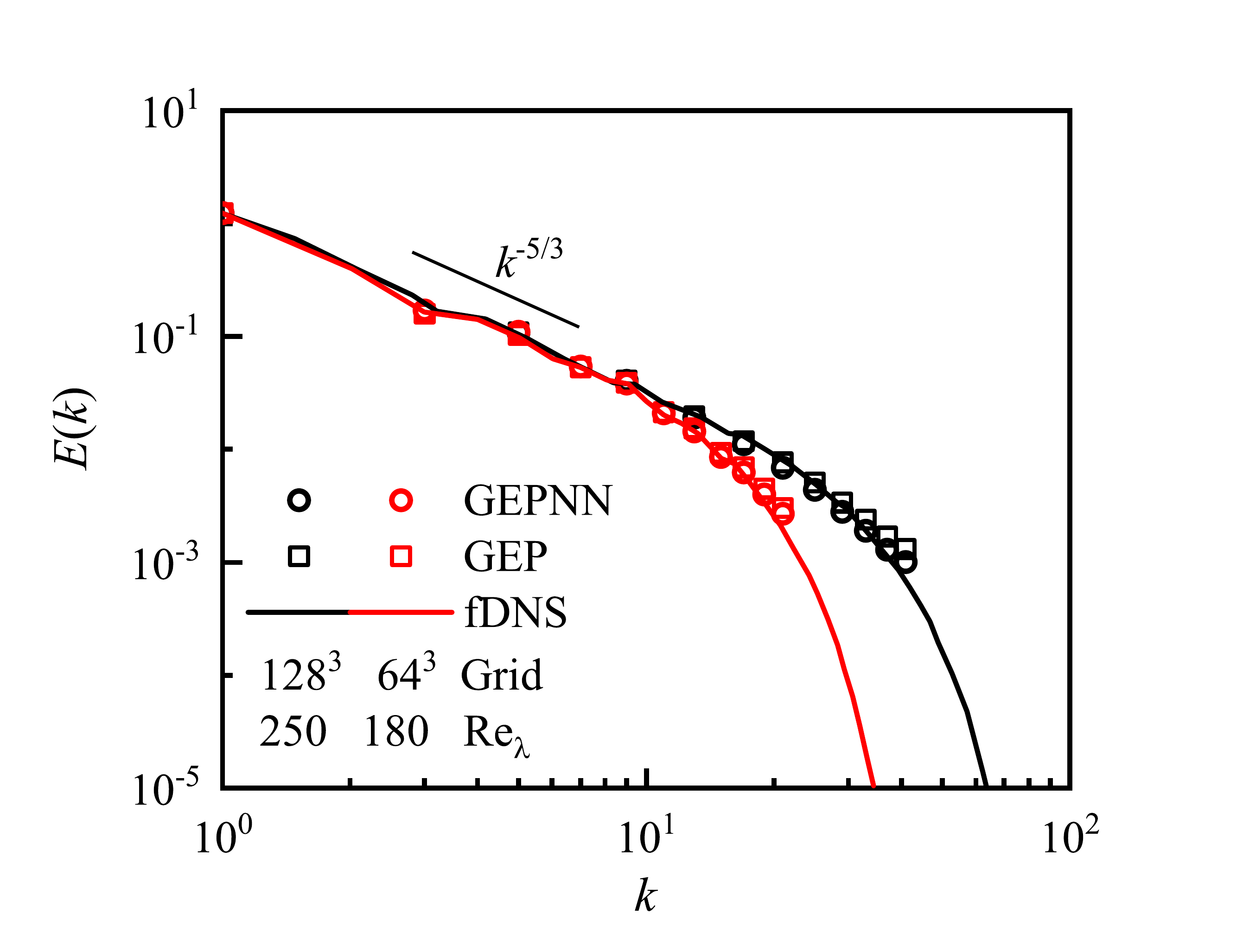}
\caption{Velocity spectrum for fDNS and corresponding LES with different Reynolds numbers and grid resolutions.}
\label{fig:sgs_general}
\end{figure}

The Reynolds number is a dimensionless variable defined as the ratio of inertial force to viscous force, and a high Reynolds number indicates that the influence of inertia is significant. 
As the flow properties vary across different Reynolds numbers, the generalization ability of the SGS model at different Reynolds numbers is critical. The spectrum of the velocity fields obtained from LES at different Reynolds numbers and grid resolutions are presented in Fig.~\ref{fig:sgs_general}. This shows that data-driven models with functional expressions have excellent generalization ability, allowing them to adapt to different flow conditions.

In the SGS stress modelling problem, the GEPNN showed better convergence performance than traditional GEP and SRNN methods. 
In particular, using a simpler structure, SRNN2 failed to model the SGS stress, and all the training sessions of SRNN2 diverged. This phenomenon reflects that the preset structure of SRNN has a strong influence on performance. Ultimately, a reliable structure needs more experience and analysis for the modelling problem, which becomes an obstacle for applying SRNN in unknown fields. The GEPNN model achieved the lowest RMSE and highest correlation coefficient in the \emph{a priori} test, which shows that the predictions of the GEPNN model correlated more closely with the DNS data. In the \emph{a posteriori} test, LES using the GEPNN model could predict turbulence statistics that align with the filtered DNS data. Compared to the traditional SGS models, the turbulence structures predicted using the GEPNN model are much finer. Moreover, the GEPNN model has a strong generalization ability, which means that the algebraic model trained from one flow condition can perform well for other Reynolds numbers and grid resolutions. 


\section{\label{sec:conc}Conclusions}
In the present study, we propose a novel method called GEPNN that aims to efficiently and robustly develop accurate models with explicit expressions via symbolic regression.
By introducing carefully designed algorithms, model expressions can be freely transformed between the GEP and SRNN methods.
In this way, the GEPNN combines the GEP's ability of generating various expression structures in the global search space with the SRNN's advantage of efficient local optimization using gradient information, compensating for the drawbacks of existing GEP and SRNN methods.
In addition, several strategies, including a regularization term in the SRNN loss function and the non-coding symbols in the GEP structures, have been introduced to enhance the sparsity of trained models, improving the model interpretability and reducing the possibility of overfitting.

The GEPNN method has then been applied to three test cases for physical models and a realistic SGS modelling problem for turbulence simulations.
The results from the physical model test cases show that with the reinforcement of the SRNN, the GEPNN can find the correct expression structures and precise constants much more efficiently and robustly compared to the traditional GEP.
Furthermore, in the SGS modelling case, GEPNN shows a stronger performance than GEP and SRNN. It can generate a variety of expressions using genetic operations and create corresponding SRNN structure to apply gradient descent, which makes up for the traditional drawbacks and integrates the advantages of GEP and SRNN. Moreover, the physical models predicted by the GEPNN have a strong generalization ability, enabling its adaptation to different application conditions.

On one hand, GEP has the advantage of global searching but exhibits weaker local optimization, especially for constants. On the other hand, the SRNN is suitable for local optimization using gradient information, but the search space of a SRNN is limited by its preset architecture. 
To address these issues, GEPNN combines GEP with SRNN, showing to be more efficient and robust at converging to explicit model expressions with accurate coefficients.
This concept of combining EA with ANN for symbolic regression tasks has the potential to be applied to different fields and applications. 

\section*{Acknowledgment}
H. Li and Y. Zhao were supported by the National Natural Science Foundation of China (Grant Nos. 92152102 and 92152202), and the Advanced Jet Propulsion Innovation Center/AEAC, funding number HKCX2022-01-010.
Y. Zhao was also financially supported by the Marine S\&T Fund of Shandong Province for Pilot National Laboratory for Marine Science and Technology (Qingdao) (No. 2022QNLM010201).
R. Sandberg acknowledges support from the ARC.



\bibliographystyle{elsarticle-num} 
\bibliography{elsarticle-template-num}





\end{document}